\documentclass[twocolumn,journal]{IEEEtran}

\usepackage{amsmath,amsthm,amssymb,mathtools,amsfonts}
\usepackage{bm} 
\usepackage{dsfont} 
\usepackage{scalerel} 

\usepackage{graphicx}
\usepackage{pgfplots}
\pgfplotsset{compat=1.17} 
\usepackage{tikz}
\usepackage{psfrag} 
\usepackage{subfigure} 
\usepackage{multirow} 
\usepackage{booktabs}
\usepackage{epstopdf}

\usepackage[ruled,vlined]{algorithm2e}

\usepackage{hyperref}
\usepackage{cite} 

\newtheorem{theo}{Theorem}
\newtheorem{lem}{Lemma}

\theoremstyle{definition}

\newtheorem{rmk}{Remark}

\usetikzlibrary{svg.path}

\definecolor{orcidlogocol}{HTML}{A6CE39}
\tikzset{
  orcidlogo/.pic={
    \fill[orcidlogocol] svg{M256,128c0,70.7-57.3,128-128,128C57.3,256,0,198.7,0,128C0,57.3,57.3,0,128,0C198.7,0,256,57.3,256,128z};
    \fill[white] svg{M86.3,186.2H70.9V79.1h15.4v48.4V186.2z}
                 svg{M108.9,79.1h41.6c39.6,0,57,28.3,57,53.6c0,27.5-21.5,53.6-56.8,53.6h-41.8V79.1z M124.3,172.4h24.5c34.9,0,42.9-26.5,42.9-39.7c0-21.5-13.7-39.7-43.7-39.7h-23.7V172.4z}
                 svg{M88.7,56.8c0,5.5-4.5,10.1-10.1,10.1c-5.6,0-10.1-4.6-10.1-10.1c0-5.6,4.5-10.1,10.1-10.1C84.2,46.7,88.7,51.3,88.7,56.8z};
  }
}
\newcommand\orcidicon[1]{\href{https://orcid.org/#1}{\mbox{\scalerel*{
\begin{tikzpicture}[yscale=-1,transform shape]
\pic{orcidlogo};
\end{tikzpicture}
}{|}}}}

\title{IRS-Assisted IoT Activity Detection Under Asynchronous Transmission and Heterogeneous Powers: Detectors and Performance Analysis}

\author{
Amirhossein Taherpour \orcidicon{0000-0003-4647-102X},~\IEEEmembership{Member,~IEEE},  
Somayeh Khani \orcidicon{0000-0003-4647-102X},~\IEEEmembership{Member,~IEEE},  
Abbas Taherpour \orcidicon{0000-0003-0706-5774},~\IEEEmembership{Senior Member,~IEEE}, and  
Tamer Khattab \orcidicon{0000-0003-2347-9555},~\IEEEmembership{Senior Member,~IEEE}  
\thanks{Amirhossein Taherpour is with the Department of Electrical Engineering, Columbia University, New York, NY, USA (e-mail: at3532@columbia.edu).  
Somayeh Khani is with the Department of Electrical Engineering, Iran University of Science and Technology, Tehran, Iran (e-mail: s.khani@elec.iust.ac.ir).  
Abbas Taherpour is with the Department of Electrical Engineering, Imam Khomeini International University, Qazvin, Iran (e-mail: abbas.taherpour@ikiu.ac.ir).  
Tamer Khattab is with the Department of Electrical Engineering, Qatar University, Doha, Qatar (e-mail: tkhattab@qu.edu.qa).}
}

\begin{document}
\maketitle
\thispagestyle{empty}
\date{}
\begin{abstract}
This paper addresses the problem of activity detection in distributed Internet of Things (IoT) networks, where devices employ asynchronous transmissions with heterogeneous power levels to report their local observations. The system leverages an intelligent reflecting surface (IRS) to enhance detection reliability, with optional incorporation of a direct line-of-sight (LoS) path. We formulate the detection problem as a binary hypothesis test and develop four detectors: an optimal detector alongside three computationally efficient detectors designed for practical scenarios with different levels of prior knowledge about noise variance, channel state information, and device transmit powers. For each detector, we derive closed-form expressions for both detection and false alarm probabilities, establishing theoretical performance benchmarks. Extensive simulations validate our analytical results and systematically evaluate the impact of key system parameters including the number of antennas, samples, users, and IRS elements on detection performance. The proposed framework effectively bridges theoretical optimality with implementation practicality, providing a scalable solution for IRS-assisted IoT networks in emerging 6G systems.
\end{abstract}
\begin{IEEEkeywords}
Activity detection, asynchronous transmission, grant-free NOMA, Internet of Things (IoT), complex Rao test, Nuisance parameters, intelligent reflecting surfaces (IRS), 6G networks.
\end{IEEEkeywords}
\section{Introduction} \label{sec:intro}

\IEEEPARstart{T}{he} rapid proliferation of the Internet of Things (IoT) has brought about a transformative shift in how physical environments are monitored, controlled, and optimized. Applications span a wide range of domains, including smart cities, industrial automation, healthcare monitoring, and environmental sensing. To support this evolution, wireless communication systems must provide unprecedented levels of scalability, reliability, and energy efficiency. As IoT networks continue to expand in size and complexity, ensuring robust connectivity and accurate device activity detection has emerged as a critical challenge, especially in dense deployments characterized by dynamic propagation environments and limited energy resources~\cite{Guo2021}.\par
A key practical barrier lies in the energy-constrained nature of IoT devices, which limits their ability to engage in frequent coordination or maintain persistent links with base stations (BSs) or access points (APs). These constraints are further exacerbated in scenarios lacking line-of-sight (LoS) paths between devices and APs, leading to unreliable and intermittent connectivity. In traditional wireless architectures, APs serve as centralized controllers and typically assume synchronized device access and uniform transmission power. However, these assumptions often fail in real-world deployments, particularly in dense urban, industrial, and indoor environments where LoS links are frequently obstructed and devices possess heterogeneous hardware capabilities and power budgets~\cite{Qamar2024,Ullah2022}.\par
Adding to this complexity is the asynchronous and sporadic transmission behavior inherent in large-scale IoT systems. Devices typically transmit a pre-defined signal vector at irregular intervals, triggered by localized sensing events, external stimuli, or available energy. This sporadic activity arises from heterogeneity in sensing modalities, battery capacities, and operational duty cycles. Consequently, transmissions are naturally uncoordinated and bursty, posing significant challenges to conventional synchronous access protocols~\cite{chen2020robust}.\par
Conventional grant-based access mechanisms, widely adopted in mobile broadband systems, are ill-suited to the sporadic and unpredictable traffic patterns typical of IoT applications. These mechanisms require devices to request transmission permission before sending data, resulting in substantial signaling overhead and increased access delays. As network density grows, the likelihood of control channel congestion, packet loss, and latency increases significantly, which poses serious challenges for time-sensitive IoT scenarios~\cite{choi2021}.\par
To address these limitations, grant-free access schemes have garnered considerable attention. By enabling devices to transmit data immediately upon detecting an event, without prior handshaking or scheduling, grant-free access offers a low-latency, low-overhead solution that naturally aligns with the asynchronous and event-driven nature of IoT networks
~\cite{shahab2020}.

Within this context, grant-free non-orthogonal multiple access (GF-NOMA) has recently been proposed to facilitate efficient data communication between heterogeneous users, such as IoT devices and BSs or APs~\cite{Yuan2021,Shaikh2024,Huang2025}. In GF-NOMA, users transmit signals at different power levels based on their energy constraints and hardware capabilities, which can also be leveraged for activity detection in IoT networks.

In practice, activity detection becomes significantly more challenging in environments where favorable propagation paths are obstructed, such as underground facilities, dense urban areas, or metallic industrial settings. In such scenarios, the weak and sporadic signals from low-power IoT devices may not be reliably captured by the AP. To address these limitations, intelligent reflecting surfaces (IRS) have emerged as a promising solution for enhancing connectivity. An IRS comprises large arrays of passive reflecting elements capable of dynamically adjusting the phase, and in some implementations the amplitude, of incoming signals to manipulate the wireless propagation environment in real time~\cite{Zou2023,Chen2022,Sun2023}. By introducing virtual LoS paths and constructively reinforcing signal reception at the AP, IRS can significantly boost both the energy efficiency and reliability of activity detection, without imposing additional power demands on the devices.

The passive architecture of IRS makes it particularly well-suited for energy-sensitive IoT applications. Moreover, its reconfigurability enables dynamic adaptation to channel variations and interference, thereby supporting more robust detection of asynchronous and low-power transmissions. The integration of IRS with grant-free access thus yields a hybrid architecture that not only enhances spatial signal diversity but also mitigates the adverse effects of transmission asynchrony and power heterogeneity, which are two key challenges in realistic IoT deployments.

\subsection{Related Work}

Early efforts on GF-NOMA for massive IoT focused on leveraging user sparsity and the high dimensionality of antenna arrays through compressive sensing techniques. These approaches enabled reliable activity detection with minimal miss detection and false alarm probabilities~\cite{Liu2018TSP}. Extensions to multi-cell networks incorporated joint estimation techniques to suppress inter-cell interference~\cite{Jiang2022TWC}, while data-driven detection strategies removed the dependency on prior activation statistics by learning optimal thresholds directly from data~\cite{Mehrabi2023CNN,shahab2023}.

To address asynchronous transmissions, which are inherent in event-driven and energy-constrained IoT environments, chaos-based pilot sequences combined with iterative sparse recovery methods have been proposed, offering robustness against unknown delays~\cite{Qiu2024WNet}. More recent developments have exploited timing offsets constructively by modeling them via factor graphs and applying sparse Bayesian learning for enhanced interference suppression~\cite{Zhang2023Arxiv}.

The integration of IRS into GF-NOMA systems has unlocked new capabilities in connectivity enhancement. Message passing algorithms have been extended to jointly detect device activity over cascaded IRS-assisted channels~\cite{Xia2021TSP}, and phase-sweeping tensor-based techniques have improved performance in unsourced random access scenarios~\cite{Shao2021GLOBECOM,Croisfelt2022Arxiv}. Under statistical channel state information (CSI), robust phase configurations optimized via the generalized likelihood ratio test (GLRT) have shown improved worst-case detection reliability~\cite{Laue2023TWC}. To further strengthen resilience, covariance-based hypothesis testing has been introduced to counteract hardware impairments~\cite{Leyva2024TWC}, while deep unfolding architectures offer data-driven alternatives with high adaptability~\cite{Zheng2024TWC,Ren2025Arxiv}.

Beyond physical-layer signal processing, asynchronous federated learning paradigms have been proposed to allow distributed IoT devices to update local models without global synchronization, substantially reducing communication overhead~\cite{Yan2023}. In latency-sensitive applications, UAV-assisted BSs have enabled flexible and energy-efficient uplink support for GF-NOMA~\cite{Fu2022}, while joint power and spectrum allocation frameworks have been developed to meet stringent quality-of-service (QoS) constraints~\cite{yahya2022power}.

Statistical detection theory continues to underpin many recent advances in GF-NOMA. The GLRT remains a widely used tool, especially within approximate message passing frameworks~\cite{Liu2018TSP,Laue2023TWC}, while generalized Rao tests offer lower-complexity alternatives under multiplicative fading~\cite{Ciuonzo2021}. For security-critical deployments, sequential multi-hypothesis testing methods have been developed to mitigate threats such as unauthorized access and jamming~\cite{Upadhyaya2021}. Collectively, these lines of work target the core challenges of asynchronous access, power heterogeneity, interference resilience, and energy efficiency, all of which this paper addresses within a unified IRS-enhanced GF-NOMA detection framework.

\subsection{Motivation and Contribution}

Existing IoT activity detection methods often rely on unrealistic assumptions, such as perfect synchronization, equal received power across devices, and rich scattering environments with predictable propagation. In contrast, practical IoT networks are characterized by asynchronous traffic with multi-symbol timing offsets, significant disparities in received power, and prevalent non-line-of-sight (NLoS) conditions. These impairments severely degrade the performance of traditional detection methods, including approximate message passing and machine learning approaches, especially in energy-constrained settings.

Although IRS offer a means to enhance signal reliability and coverage, current IRS-assisted detection strategies are limited. Many assume perfect CSI, fixed IRS configurations, or involve substantial training overhead. Furthermore, existing schemes typically address only isolated issues such as asynchrony or power imbalance, but not their joint effects under practical constraints.

This paper addresses the need for robust and scalable activity detection in uncoordinated, low-power IoT networks by investigating an IRS-assisted uplink scenario that incorporates asynchronous device transmissions and heterogeneous power levels. The objective is to design low-complexity detection strategies that utilize passive IRS beamforming to mitigate unfavorable channel conditions, reinforce weak signals, and maintain detection robustness under timing and power disparities. Our analysis captures the joint effects of random device activity, asynchrony, and partial channel knowledge, highlighting the role of adaptive IRS deployment in enabling reliable detection with minimal coordination.

We propose an activity detection framework tailored for the requirements of 6G IoT environments, where latency sensitivity, limited energy resources, and uncertain noise levels are critical challenges. In our system model, each IoT device transmits an activity signal with a power level reflecting its energy budget. These signals are passively reflected by the IRS toward an AP, and when possible, a direct device-to-AP LoS path is also utilized to improve detection diversity.

Four detectors are developed, addressing varying levels of system uncertainty. These include an optimal detector assuming full knowledge of all parameters; a detector robust to unknown noise variance; a detector for scenarios with unknown transmit powers and channel coefficients; and two detectors for the most general and realistic case where noise variance, power levels, and signal waveforms are all unknown. For this challenging case, we introduce a GLRT and a Complex Rao-based test, both designed to operate effectively under parameter uncertainty while exploiting the reflective properties of IRS. We derive the probability of detection and false alarm analytically for all of these detectors and analyze the impact of different system parameters on the activity detection.

The rest of this paper is organized as follows. Section~\ref{sec:model} describes the problem formulation and basic assumptions, as well as the binary hypothesis testing framework. Section~\ref{sec:OP} derives the optimal Neyman–Pearson detector under full knowledge of the system parameters. In Section~\ref{sec:Sub-OP}, three low-complexity sub-optimal detectors are developed: two GLRT-based tests for special cases of unknown parameters and a Complex Rao test for the fully blind scenario. Section~\ref{sec:Per} provides closed-form, asymptotic expressions for the detection and false-alarm probabilities of all proposed detectors. Section~\ref{subsec:Per_Sub_OP2} analyzes the case of unknown device transmit powers, and Section~\ref{subsec:parameter_impact} examines the impact of key system parameters on performance and derives practical design guidelines. Simulation results and in-depth discussions are presented in Section~\ref{sec:simulation}, and Section~\ref{sec:conclusion} concludes the paper.

\subsection{Notation}
\label{sec:notation}

The lightface letters denote scalars (e.g., $x$, $\sigma^2$) while boldface lowercase and uppercase letters represent column vectors (e.g., $\boldsymbol{v}$, $\boldsymbol{h}_i$) and matrices (e.g., $\boldsymbol{H}$, $\boldsymbol{\Sigma}_n$) respectively. The all-zero and all-one vectors are denoted by $\boldsymbol{0}$ and $\boldsymbol{1}$, with $\boldsymbol{O}$ representing the all-zero matrix and $\boldsymbol{I}$ the identity matrix. Diagonal matrices are expressed as $\boldsymbol{P} = \mathrm{diag}\{p_{1},...,p_{K}\}$. Matrix operations use superscripts $*$, $T$, $H$, $\dagger$ for complex conjugate, transpose, Hermitian transpose, and Moore-Penrose pseudo-inverse respectively, while $\hat{(\cdot)}$ indicates estimation. The Frobenius norm is denoted $\|\boldsymbol{A}\|_F$, with $\mathrm{tr}(\boldsymbol{B})$ and $\mathrm{etr}(\boldsymbol{B})$ representing the trace and exponential trace ($e^{\mathrm{tr}(\boldsymbol{B})}$) of matrix $\boldsymbol{B}$. Order notation follows standard asymptotic analysis: $f(x) = \mathcal{O}(g(x))$ denotes $|f(x)| \leq C|g(x)|$ for some constant $C > 0$ as $x \to \infty$, and $f(x) = o(g(x))$ indicates $f(x)/g(x) \to 0$. The imaginary unit $j$ satisfies $j^2 = -1$, with $\Re(\boldsymbol{c})$ and $\Im(\boldsymbol{c})$ denoting real and imaginary parts of complex vector $\boldsymbol{c}$. The statistical expectation operator is $\mathbb{E}\{\cdot\}$, and $\mathbb{C}$ represents the field of complex numbers. Matrix sets include $\mathbb{C}^{n \times n}_{S}$ (positive semidefinite) and $\mathbb{C}^{n \times n}_{+}$ (positive definite). The $\mathcal{CN}(\boldsymbol{\mu}, \boldsymbol{\Sigma})$ denotes the complex Gaussian distribution with mean vector $\boldsymbol{\mu}$ and covariance matrix $\boldsymbol{\Sigma}$.
\section{Problem Statement and Basic Assumptions}\label{sec:model} 
\begin{figure}[t!]
    \centering
\includegraphics[width=0.9\linewidth]{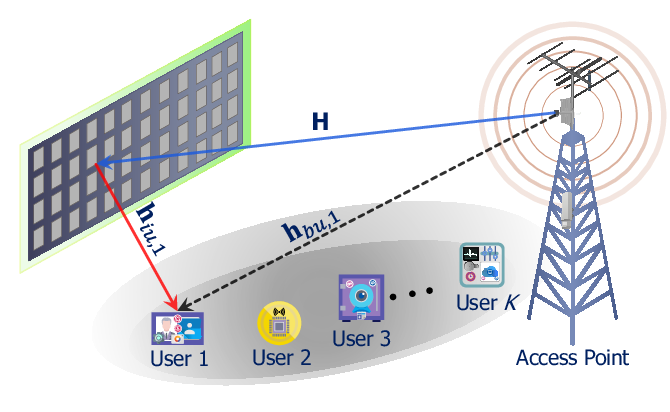} 
    \caption{System model for asynchronous IoT devices.}
    \label{sysmodel}
\end{figure}

We consider an IRS-aided uplink system comprising a BS/AP equipped with \( M \) antennas and \( K \) IoT devices as shown in Fig.~\ref{sysmodel}. Each device is responsible for sensing and transmitting environmental information such as traffic, air quality, energy usage, or patient vitals. The transmission occurs in an asynchronous and grant-free manner, where devices transmit their signals without coordination or scheduling. 

To handle multiple devices transmitting simultaneously, GF-NOMA is employed, enabling signal separation based on power control rather than time or frequency division. Each device \( i \in \mathcal{K} = \{1, 2, \dots, K\} \) transmits with power \( p_i \), determined by its battery, range, and practical capabilities. The power allocation is represented by the diagonal power matrix:
\begin{equation}
\bm{P} = \mathrm{diag}\{p_1, p_2, \dots, p_K\} \in \mathbb{C}^{K \times K}. \label{eq1}
\end{equation}

The transmitted signal from the devices is subject to power scaling. Let \( \tilde{\bm{s}}_l = [\tilde{s}_{1l}, \tilde{s}_{2l}, \dots, \tilde{s}_{Kl}]^\mathrm{T} \in \mathbb{C}^{K \times 1} \) denote the original signal vector transmitted by the \( K \) devices at time \( l \). After power scaling, the transmitted signal vector for device \( i \) at time \( l \) becomes:
\begin{equation}
s_i[l] = \sqrt{p_i} \, \tilde{s}_i[l]. \label{eq2}
\end{equation}
Stacking the signals of all \( K \) devices, the transmitted signal vector can be written as:
\begin{equation}
\bm{s}[l] = \bm{P}^{1/2} \tilde{\bm{s}}[l] \in \mathbb{C}^{K \times 1}. \label{eq3}
\end{equation}

All devices use the IRS to transmit their signals to the BS. The IRS consists of \( N \) passive elements, and its reflection matrix is given by:
\begin{equation}
\bm{\Phi} = \mathrm{diag}(e^{j\theta_1}, e^{j\theta_2}, \dots, e^{j\theta_N}) \in \mathbb{C}^{N \times N}, \label{eq4}
\end{equation}
where \( \theta_n \) represents the phase shift of the \( n \)-th IRS element (assuming ideal reflection with unit amplitude).

The channel model incorporates both IRS-aided and potential direct paths. The device-to-IRS channels are represented by the matrix \(\bm{E} = [\bm{e}_1, \bm{e}_2, \dots, \bm{e}_K] \in \mathbb{C}^{N \times K}\), where \(\bm{e}_i \in \mathbb{C}^{N}\) is the channel from device \(i\) to the IRS. The IRS-to-AP channel is denoted by \(\bm{F} = [\bm{f}_1, \bm{f}_2, \dots, \bm{f}_N] \in \mathbb{C}^{M \times N}\), where \(\bm{f}_j \in \mathbb{C}^{M}\) is the channel from the \(j\)-th IRS element to the AP. The potential direct device-to-AP channels are collected in \(\widetilde{\bm{D}} = [\bm{d}_1, \bm{d}_2, \dots, \bm{d}_K] \in \mathbb{C}^{M \times K}\), with \(\bm{d}_i \in \mathbb{C}^{M}\) representing the direct channel from device \(i\) to the AP when present. The presence of direct links is indicated by the binary vector \(\bm{\delta} = [\delta_1, \delta_2, \dots, \delta_K]^T \in \{0,1\}^K\), where \(\delta_i = 1\) if device \(i\) has a direct path to the AP.

The effective direct channel matrix is constructed as:
\begin{equation}
\bm{D} = \widetilde{\bm{D}} \odot (\bm{1}_M \bm{\delta}^T) \in \mathbb{C}^{M \times K}, \label{eq5}
\end{equation}
where \(\odot\) denotes the Hadamard product and \(\bm{1}_M\) is an all-ones vector of length \(M\).

The composite IRS-assisted channel for all devices is:
\begin{equation}
\bm{H}^{\text{IRS}} = \bm{F} \bm{\Phi} \bm{E} \in \mathbb{C}^{M \times K}. \label{eq6}
\end{equation}

The total effective channel matrix combining both IRS-reflected and direct paths is:
\begin{equation}
\bm{H} = \bm{H}^{\text{IRS}} + \bm{D} = \bm{F} \bm{\Phi} \bm{E} + \widetilde{\bm{D}} \odot (\bm{1}_M \bm{\delta}^T) \in \mathbb{C}^{M \times K}. \label{eq7}
\end{equation}

At the BS, due to the asynchronous nature of the system, each signal experiences an unknown delay \( \tau_i \) before arriving. The BS collects \( L \) samples of the received signal over an observation interval \( T \), with a sampling frequency \( f_s \). To ensure all delayed signals are fully captured, the observation interval must satisfy:
\begin{equation}
\min_{i \in \mathcal{K}} \tau_i \leq T \quad \text{and} \quad T + L/f_s \geq \max_{i \in \mathcal{K}} \tau_i. \label{eq8}
\end{equation}
Here, \( \min_{i \in \mathcal{K}} \tau_i \) and \( \max_{i \in \mathcal{K}} \tau_i \) represent the minimum and maximum delays among all devices, respectively. In practice, the BS does not know the exact delays \( \tau_i \) in advance. However, by assuming a sufficiently large observation window \( T \) and sampling duration \( L/f_s \), the BS can ensure that all signals, regardless of their delays, are captured within the sampling window. This assumption is justified in scenarios where the range of possible delays is known or can be estimated based on the deployment environment.

At each time \( l \), the received signal at the \( M \) antennas of the BS is given by:
\begin{equation}
\bm{x}[l] = \bm{H} \bm{P}^{1/2} \bm{s} + \bm{n}[l], \quad l = 1, 2, \dots, L, \label{eq9}
\end{equation}
where \( \bm{x}[l] \in \mathbb{C}^{M \times 1} \) is the received signal vector at time \( l \), \( \bm{s} \in \mathbb{C}^{K \times 1} \) is the signal vector transmitted by the \( K \) devices, and \( \bm{n}[l] \in \mathbb{C}^{M \times 1} \) is the noise vector at time \( l \).
The noise at the \( l \)-th observation is modeled as a zero-mean circularly symmetric complex Gaussian vector:
\begin{equation}
\bm{n}[l] \sim \mathcal{CN}_M(\bm{0}_{M \times 1}, \bm{\Sigma}_n), \label{eq10}
\end{equation}
where \( \bm{\Sigma}_n \in \mathbb{C}^{M \times M}_{+} \) is the noise covariance matrix.

 For ease of notation and compactness, from now on, we use \( \bm{x}_l \) and \( \bm{n}_l \) interchangeably with \( \bm{x}[l] \) and \( \bm{n}[l] \), respectively.

Now, the received signal at the BS is modeled under two hypotheses:
\begin{equation}
\begin{cases}
\mathcal{H}_0: & \bm{x}_l = \bm{n}_l, \\
\mathcal{H}_1: & \bm{x}_l = \bm{H} \bm{P}^{1/2} \bm{s} + \bm{n}_l,
\end{cases} \label{eq11}
\end{equation}
where \( \bm{x}_l \in \mathbb{C}^{M \times 1} \) is the \( l \)-th received sample vector, and \( \bm{n}[l] \sim \mathcal{CN}_M(\bm{0}_{M \times 1}, \bm{\Sigma}_n) \) is the noise vector at time \( l \).

The BS collects \( L \) samples, which are stacked into the matrix:
\begin{equation}
\bm{X} = [\bm{x}_1, \bm{x}_2, \dots, \bm{x}_L] \in \mathbb{C}^{M \times L}. \label{eq12}
\end{equation}
The received signal model is compactly expressed as:
\begin{equation}
\bm{X} = \bm{H} \bm{P}^{1/2} \bm{S} + \bm{N}, \label{eq13}
\end{equation}
where \( \bm{S} = \bm{s} \bm{1}^\mathrm{T} \in \mathbb{C}^{K \times L} \) is the transmitted signal matrix, \( \bm{1} \in \mathbb{C}^{L} \) is an all-ones vector, and \( \bm{N} = [\bm{n}_1, \bm{n}_2, \dots, \bm{n}_L] \in \mathbb{C}^{M \times L} \) is the additive noise matrix with i.i.d. entries following \( \mathcal{CN}_M(\bm{0}_{M \times 1}, \bm{\Sigma}_n) \). Now the hypothesis testing can be written as:

\begin{equation}
\begin{cases}
\mathcal{H}_0: & \bm{X} = \bm{N}, \\
\mathcal{H}_1: & \bm{X} = \bm{H} \bm{P}^{1/2} \bm{S} + \bm{N},
\end{cases} \label{eq14S}
\end{equation}
\section{Optimum Detector}\label{sec:OP}
As an ideal scenario, when the BS has access to the complete set of information, including the effective channel gain matrix $ \boldsymbol{H} $, noise covariance matrix $ \boldsymbol{\Sigma}_n $, and transmit power matrix $ \boldsymbol{P} $, the optimal detector based on the Neyman-Pearson (NP) criterion can be derived for comparison and used as a performance benchmark.

From \ref{eq14S}, the probability density function (PDF) of the observation matrix \( \boldsymbol{X} \) under the two hypotheses is given by:
\begin{align}
f(\boldsymbol{X}|\mathcal{H}_0) &= (\pi)^{-ML} \, |\boldsymbol{\Sigma}_n|^{-L} \, \text{etr}\left\{-\boldsymbol{\Sigma}_n^{-1} \boldsymbol{X} \boldsymbol{X}^H \right\}, \label{eq14} \\
f(\boldsymbol{X}|\mathcal{H}_1) &= (\pi)^{-ML} \, |\boldsymbol{\Sigma}_n|^{-L} \label{eq15} \\
&\quad \times \text{etr}\left\{-\boldsymbol{\Sigma}_n^{-1} (\boldsymbol{X} - \boldsymbol{H} \boldsymbol{P}^{1/2} \boldsymbol{S}) (\boldsymbol{X} - \boldsymbol{H} \boldsymbol{P}^{1/2} \boldsymbol{S})^H \right\}. 
\nonumber
\end{align}
Here, \( \text{etr}(\cdot) \) denotes the exponential trace function.

The likelihood ratio (LR) function is given by:
\begin{align}\label{e5-01}
LR(\boldsymbol{X}) &= \frac{f(\boldsymbol{X}|\mathcal{H}_1)}{f(\boldsymbol{X}|\mathcal{H}_0)}  \\
&= \text{etr} \big\{ \boldsymbol{\Sigma}_n^{-1} \big( 
\boldsymbol{H} \boldsymbol{P}^{1/2} \boldsymbol{S} \boldsymbol{X}^H 
+ \boldsymbol{X} \boldsymbol{S}^H \boldsymbol{P}^{1/2} \boldsymbol{H}^H \nonumber \\
&\qquad\qquad\quad
- \boldsymbol{H} \boldsymbol{P}^{1/2} \boldsymbol{S} \boldsymbol{S}^H \boldsymbol{P}^{1/2} \boldsymbol{H}^H 
\big) \big\}
\nonumber
\end{align}

By using the logarithm of the likelihood ratio (LLR) and discarding common and constant terms, the optimal test statistic is obtained as:
\begin{align}
\Lambda(\boldsymbol{X}) = \operatorname{tr} \left\{ \boldsymbol{\Sigma}_n^{-1} \left( \boldsymbol{H} \boldsymbol{P}^{1/2} \boldsymbol{S} \boldsymbol{X}^H + \boldsymbol{X} \boldsymbol{S}^H \boldsymbol{P}^{1/2} \boldsymbol{H}^H \right) \right\} \mathop{\gtrless}_{\mathcal{H}_0}^{\mathcal{H}_1} \tau \label{e5-02}
\end{align}

Based on the above expression, the decision statistic for the optimal NP detector can be simplified as:
\begin{align}\label{e5} 
T_{\rm Opt}(\boldsymbol{X}) =
2 \Re \left\{ \operatorname{tr} \left\{ \boldsymbol{M}^H \boldsymbol{\Sigma}_n^{-1} \boldsymbol{X} \right\} \right\} \mathop{\gtrless}_{\mathcal{H}_0}^{\mathcal{H}_1} \tau_{\rm NP}
\end{align}
where $ \boldsymbol{M} \triangleq \boldsymbol{H} \boldsymbol{P}^{1/2} \boldsymbol{S} $.
\section{Sub-Optimum Detectors}\label{sec:Sub-OP}
When some of the parameters are unknown, this lack of knowledge results to a reduction in the detection performance in comparison with an optimum detector and we have to solve a composite hypothesis testing problem.

One of the most commonly used approaches for solution of the composite hypothesis testing problems is GLRT. In this method, the maximum likelihood estimation (MLE) of the unknown parameters are used to construct the likelihood ratio function. MLE is optimal for large data size, asymptotically unbiased and efficient.  

In this section, we consider three special cases which all of the other practical scenarios can be transformed and expressed as one of these three cases:
\begin{enumerate} 
\item Noise covariance matrix $ \boldsymbol{{\Sigma}_n} $ is unknown at the BS.
\item Power matrix $ \boldsymbol{P}$ and transmitted signals matrix from devices $ \boldsymbol{S} $ are unknown at the BS.
\item Channel gain matrix $ \boldsymbol{H} $, $ \boldsymbol{{\Sigma}_n} $, $ \boldsymbol{P} $, and $ \boldsymbol{S} $ are unknown together.
\end{enumerate}

\subsection{Unknown Noise}

In this case, we assume the covariance matrix of noise is unknown and derive the GLRT. The PDF of the received data matrix $\boldsymbol{X} \in \mathbb{C}^{M \times L}$ under hypotheses $\mathcal{H}_0$ and $\mathcal{H}_1$ are given by \eqref{eq14} and \eqref{eq15}.

Since the Gaussian distribution belongs to the exponential family, and using the Neyman-Fisher factorization theorem, it follows that the sample covariance matrix is a complete sufficient statistic for estimating the unknown covariance matrix. Furthermore, this estimator is unbiased, making it the minimum variance unbiased estimator (MVUE) \cite{r54}. Therefore, the sample covariance matrices under $\mathcal{H}_0$ and $\mathcal{H}_1$ are:
\begin{align}
\hat{\boldsymbol{\Sigma}}_n|\mathcal{H}_0 &= \frac{1}{L} \boldsymbol{X} \boldsymbol{X}^H, \label{e6}\\
\hat{\boldsymbol{\Sigma}}_n|\mathcal{H}_1 &= \frac{1}{L}(\boldsymbol{X} - \boldsymbol{H} \boldsymbol{P}^{1/2} \boldsymbol{S})(\boldsymbol{X} - \boldsymbol{H} \boldsymbol{P}^{1/2} \boldsymbol{S})^H. \label{e7}
\end{align}

We assume that the noise is spatially white and identically distributed across BS antennas, i.e., $\boldsymbol{\Sigma}_n = \sigma^2 \boldsymbol{I}_M$. Substituting this into \eqref{e6} and \eqref{e7}, the maximum likelihood estimates of the noise power under both hypotheses become:
\begin{align}
\hat{\sigma}_0^2 &= \frac{1}{ML} \text{tr}\left\{\boldsymbol{X} \boldsymbol{X}^H \right\}, \label{e8} \\
\hat{\sigma}_1^2 &= \frac{1}{ML} \text{tr}\left\{ (\boldsymbol{X} - \boldsymbol{H} \boldsymbol{P}^{1/2} \boldsymbol{S})(\boldsymbol{X} - \boldsymbol{H} \boldsymbol{P}^{1/2} \boldsymbol{S})^H \right\}. \label{e9}
\end{align}

The GLRT decision rule is then defined as the ratio of the maximized likelihoods under $\mathcal{H}_1$ and $\mathcal{H}_0$. 
Taking the logarithm and simplifying using the determinant and trace properties, we get:
\begin{equation}
\ln \Lambda_{GLR}(\boldsymbol{X}) = ML \ln \left( \frac{\hat{\sigma}_0^2}{\hat{\sigma}_1^2} \right)
\end{equation}

Equivalently, the GLRT statistic becomes:
\begin{equation}
T_{GLR}(\boldsymbol{X}) = \frac{\hat{\sigma}_0^2}{\hat{\sigma}_1^2} = \frac{\text{tr}\left\{ \boldsymbol{X} \boldsymbol{X}^H \right\}}{\text{tr}\left\{ (\boldsymbol{X} - \boldsymbol{H} \boldsymbol{P}^{1/2} \boldsymbol{S})(\boldsymbol{X} - \boldsymbol{H} \boldsymbol{P}^{1/2} \boldsymbol{S})^H \right\}},
\end{equation}

By further simplifying the numerator and expanding the quadratic term in the denominator, we obtain:
\begin{align}
&\text{tr}\left\{ (\boldsymbol{X} - \boldsymbol{M})(\boldsymbol{X} - \boldsymbol{M})^H \right\} \nonumber \\
&\quad= \text{tr}\left\{ \boldsymbol{X}\boldsymbol{X}^H - \boldsymbol{M} \boldsymbol{X}^H - \boldsymbol{X} \boldsymbol{M}^H + \boldsymbol{M} \boldsymbol{M}^H \right\} \nonumber \\
&\quad= \text{tr}\left\{ \boldsymbol{X}\boldsymbol{X}^H \right\} 
- 2\Re\left\{ \text{tr}(\boldsymbol{M}^H \boldsymbol{X}) \right\}
+ \text{tr}\left\{ \boldsymbol{M}^H \boldsymbol{M} \right\}.
\end{align}

where as defined $\boldsymbol{M} = \boldsymbol{H} \boldsymbol{P}^{1/2} \boldsymbol{S}$. Substituting into the GLRT expression gives the final test statistic:
\begin{equation} \label{e10}
T_2(\boldsymbol{X}) = \frac{\text{tr}\left\{ 2 \Re\left\{ \boldsymbol{M}^H \boldsymbol{X} \right\} - \boldsymbol{M}^H \boldsymbol{M} \right\}}{\text{tr}\left\{ \boldsymbol{X}^H \boldsymbol{X} \right\}} \mathop{\gtrless}_{\mathcal{H}_0}^{\mathcal{H}_1} \tau_{\rm NP}.
\end{equation}

\subsubsection{Low-SNR Approximation}
When the signal power is significantly weaker than the noise power ($\|\bm{M}\|_F^2 \ll \sigma^2$), the GLRT statistic in \eqref{e10} can be simplified through careful analysis of its constituent terms. Beginning with the denominator $\text{tr}(\bm{X}^H \bm{X})$, we observe that under $\mathcal{H}_1$ where $\bm{X} = \bm{M} + \bm{N}$, the expansion yields:

\begin{equation}
\text{tr}(\bm{X}^H \bm{X}) = \|\bm{M}\|_F^2 + 2\Re\{\text{tr}(\bm{M}^H \bm{N})\} + \|\bm{N}\|_F^2.
\end{equation}

In the low-SNR regime, the noise-dominated term $\|\bm{N}\|_F^2$ becomes the principal component, allowing the approximation:

\begin{equation}
\text{tr}(\bm{X}^H \bm{X}) \approx \|\bm{N}\|_F^2 + \mathcal{O}(\|\bm{M}\|_F).
\end{equation}

The numerator $2\Re\{\text{tr}(\bm{M}^H \bm{X})\} - \|\bm{M}\|_F^2$ similarly simplifies since the quadratic signal term $\|\bm{M}\|_F^2$ becomes negligible compared to the cross-term:

\begin{equation}
2\Re\{\text{tr}(\bm{M}^H \bm{X})\} - \|\bm{M}\|_F^2 \approx 2\Re\{\text{tr}(\bm{M}^H \bm{N})\}.
\end{equation}

Consequently, the GLRT statistic reduces to a normalized matched filter:

\begin{equation}
T_{\text{Low-SNR}}(\bm{X}) = \frac{\Re\{\text{tr}(\bm{M}^H \bm{X})\}}{\text{tr}(\bm{X}^H \bm{X})} 
\mathop{\gtrless}_{\mathcal{H}_0}^{\mathcal{H}_1} \tau_{\rm NP}',
\label{eq:low_snr_stat}
\end{equation}

where $\tau_{\rm NP}'$ absorbs all constant terms into the modified detection threshold.

\subsubsection{Large-Sample ($L \gg 1$) Approximation}
For scenarios with abundant samples ($L \to \infty$), the sample covariance matrix converges to its statistical expectation. The denominator's asymptotic behavior follows from:

\begin{equation}
\frac{1}{L}\text{tr}(\bm{X}^H \bm{X}) \to \text{tr}(\bm{\Sigma}_n) + \frac{1}{L}\|\bm{M}\|_F^2 = M\sigma^2 + \frac{1}{L}\|\bm{M}\|_F^2,
\end{equation}

yielding the approximation:

\begin{equation}
\text{tr}(\bm{X}^H \bm{X}) \approx LM\sigma^2 + \|\bm{M}\|_F^2.
\end{equation}

The numerator's limiting form emerges from the dominance of the cross-term:

\begin{equation}
2\Re\{\text{tr}(\bm{M}^H \bm{X})\} - \|\bm{M}\|_F^2 \approx \|\bm{M}\|_F^2 + 2\Re\{\text{tr}(\bm{M}^H \bm{N})\}.
\end{equation}

Substituting these into \eqref{e10} and noting that $\|\bm{M}\|_F^2$ becomes insignificant in the denominator for large $L$, we obtain:

\begin{equation}
T_2(\bm{X}) \approx \frac{2\Re\{\text{tr}(\bm{M}^H \bm{N})\}}{LM\sigma^2} + \frac{\|\bm{M}\|_F^2}{LM\sigma^2}.
\end{equation}

Dropping the constant term (which merges with the threshold) yields the large-sample detector:

\begin{equation}
T_{\text{High-}L}(\bm{X}) = \frac{\Re\{\text{tr}(\bm{M}^H \bm{X})\}}{L} 
\mathop{\gtrless}_{\mathcal{H}_0}^{\mathcal{H}_1} \tau_{\rm NP}'',
\label{eq:high_L_stat}
\end{equation}

\subsection{Unknown Channel and Power}\label{subsec:P}

When both the power profile matrix $\boldsymbol{P}$ of the IoT devices and the effective channel matrix $\boldsymbol{H}$ between the BS and the IoT devices are unknown under $\mathcal{H}_1$, we first derive the MLE for $\boldsymbol{P}$. The received signal under $\mathcal{H}_1$ is modeled as
\begin{equation}
\boldsymbol{X} = \boldsymbol{H} \boldsymbol{P}^{\frac{1}{2}} \boldsymbol{s} \boldsymbol{1}^T + \boldsymbol{N},
\end{equation}
where $\boldsymbol{s} \in \mathbb{C}^{K \times 1}$ is the known pilot signal, $\boldsymbol{P}^{\frac{1}{2}} \in \mathbb{R}^{K \times K}$ is a diagonal matrix with the square roots of the IoT devices’ transmit powers, and $\boldsymbol{N} \sim \mathcal{CN}(\boldsymbol{0}, \boldsymbol{\Sigma}_n)$ is complex Gaussian noise.

The log-likelihood function under $\mathcal{H}_1$ is given by
\begin{align}
\ln f(\boldsymbol{X}|\mathcal{H}_1) &= -\mathrm{Tr}\left\{ \left( \boldsymbol{X} - \boldsymbol{H} \boldsymbol{P}^{\frac{1}{2}} \boldsymbol{s} \boldsymbol{1}^T \right)^H \boldsymbol{\Sigma}_n^{-1} \right. \notag \\
& \quad \times \left. \left( \boldsymbol{X} - \boldsymbol{H} \boldsymbol{P}^{\frac{1}{2}} \boldsymbol{s} \boldsymbol{1}^T \right) \right\} + C.
\end{align}

where $C$ is a constant independent of the unknown parameters.

To find the MLE of $\boldsymbol{P}$, we differentiate the log-likelihood function with respect to $\boldsymbol{P}^{\frac{1}{2}}$:
\begin{align}\label{e11}
\frac{\partial \ln f(\boldsymbol{X}|\mathcal{H}_1)}{\partial \boldsymbol{P}^{\frac{1}{2}}}
= 2\Re\left( \boldsymbol{s}^* \boldsymbol{1}^T \left( \boldsymbol{X} - \boldsymbol{H} \boldsymbol{P}^{\frac{1}{2}} \boldsymbol{s} \boldsymbol{1}^T \right)^T \boldsymbol{\Sigma}_n^{-T} \boldsymbol{H}^* \right).
\end{align}

By setting the gradient in \eqref{e11} to zero, we obtain the MLE of the power profile matrix as:
\begin{equation}\label{e12}
\hat{\boldsymbol{P}}^{\frac{1}{2}} = \frac{1}{L} \boldsymbol{H}^{-1}_{\rm left} \boldsymbol{X} \boldsymbol{1} \boldsymbol{s}^H \left( \boldsymbol{s} \boldsymbol{s}^H \right)^{\dagger},
\end{equation}
where $\left( \boldsymbol{s} \boldsymbol{s}^H \right)^{\dagger}$ denotes the Moore-Penrose pseudo-inverse of $\boldsymbol{s} \boldsymbol{s}^H$, which can be computed using singular value decomposition (SVD).

In the case where the IoT devices are distributed, their activities do not interfere with each other. Assuming no correlation among the IoT devices, the channel matrix $\boldsymbol{H}$ typically has full column rank. Hence, we adopt the left inverse of $\boldsymbol{H}$, defined as:
\begin{equation}
\boldsymbol{H}_{\mathrm{left}}^{-1} \triangleq \left( \boldsymbol{H}^H \boldsymbol{H} \right)^{-1} \boldsymbol{H}^H,
\end{equation}
which satisfies $\boldsymbol{H}^{-1}_{\rm left} \boldsymbol{H} = \boldsymbol{I}$. On the other hand, when there is correlation between the IoT devices or LoS dominance, the channel matrix may be rank-deficient. In such cases, the Moore-Penrose inverse must be used to ensure valid estimation.

Hence, applying the NP criterion with the MLE of $\boldsymbol{P}^{\frac{1}{2}}$ substituted back into the LR, the resulting decision statistic becomes:
\begin{equation}\label{e13}
T_{2}(\boldsymbol{X}) = \boldsymbol{1}^{T} \boldsymbol{X}^{H} \boldsymbol{\Sigma}_n^{-1} \boldsymbol{X} \boldsymbol{1} \mathop{\gtrless}_{\mathcal{H}_0}^{\mathcal{H}_1} \tau_{\rm NP},
\end{equation}
where $\tau_{\rm NP}$ is the decision threshold determined based on a desired false alarm probability level.
\subsection{Blind Detector}\label{subsec:blind}

In this section, we derive two detectors for the case where $\boldsymbol{P}$, $\boldsymbol{{\Sigma}_n}$, and $\boldsymbol{H}$ are unknown. The first detector is based on the GLRT, and the second is derived using the complex Rao test.

\subsubsection{GLRT-Based Detector}

We first define the matrix $\boldsymbol{G} \triangleq \boldsymbol{H}\boldsymbol{P}^{\frac{1}{2}} \in \mathds{C}^{M \times P}$, where the MLE of both $\boldsymbol{P}^{\frac{1}{2}}$ and $\boldsymbol{H}$ can be obtained under the hypothesis $\mathcal{H}_1$.

To estimate $\boldsymbol{G}$, we compute the derivative of the LLR with respect to $\boldsymbol{G}$, given by:
\begin{align}\label{e14}
\frac{\partial \ln{f(\boldsymbol{X}|\mathcal{H}_1)}}{\partial \boldsymbol{G}} &= \boldsymbol{{\Sigma}_n}^{-T} \left( \boldsymbol{X} - \boldsymbol{G} \boldsymbol{s} \boldsymbol{1}^T \right)^* \boldsymbol{1} \boldsymbol{s}^T \notag \\
&= \boldsymbol{{\Sigma}_n}^{-T} \left( \boldsymbol{X} - \boldsymbol{G} \boldsymbol{s} \boldsymbol{1}^T \right)^H \boldsymbol{1} \boldsymbol{s}^T.
\end{align}
The first term represents the deviation of the received signal from the expected signal model, while the second term acts as a scaling factor involving the signal vector $\boldsymbol{s}$ and the constant vector $\boldsymbol{1}$.

To estimate $\boldsymbol{G}$, we set the derivative to zero and solve for $\boldsymbol{G}$, yielding the following expression:
\begin{equation}\label{e15}
\hat{\boldsymbol{G}} = \frac{1}{L} \boldsymbol{X} \boldsymbol{1} \boldsymbol{s}^H \left( \boldsymbol{s} \boldsymbol{s}^H \right)^{\dagger}.
\end{equation}
By substituting the expression for $\hat{\boldsymbol{G}}$ back into the model $\boldsymbol{X} = \boldsymbol{G} \boldsymbol{s} \boldsymbol{1}^T + \boldsymbol{N}$ under $\mathcal{H}_1$, we obtain the ML estimate of the noise covariance matrix:
\begin{equation}\label{e17}
\hat{\boldsymbol{\Sigma}}_n^{(1)} = \frac{1}{L} \left( \boldsymbol{X} - \hat{\boldsymbol{G}} \boldsymbol{s} \boldsymbol{1}^T \right) \left( \boldsymbol{X} - \hat{\boldsymbol{G}} \boldsymbol{s} \boldsymbol{1}^T \right)^H.
\end{equation}
Under $\mathcal{H}_0$, where the signal is absent and only noise is observed, the ML estimate of the noise covariance matrix simplifies to:
\begin{equation}\label{e18}
\hat{\boldsymbol{\Sigma}}_n^{(0)} = \frac{1}{L} \boldsymbol{X} \boldsymbol{X}^H.
\end{equation}

Using the property of the pseudo-inverse, $\boldsymbol{s}^H \left( \boldsymbol{s} \boldsymbol{s}^H \right)^{\dagger} \boldsymbol{s} = 1$, the LR test statistic simplifies to:
\begin{align}\label{e16}
T_3(\boldsymbol{X}) &=  \frac{\left| \boldsymbol{X} \boldsymbol{X}^H - \frac{1}{L} \boldsymbol{X} \boldsymbol{1} \boldsymbol{1}^T \boldsymbol{X}^H \right|^{-L}}{\left| \boldsymbol{X} \boldsymbol{X}^H \right|^{-L}} \mathop{\gtrless}_{\mathcal{H}_0}^{\mathcal{H}_1} \tau_{\rm NP}.
\end{align}

\subsubsection{Complex Rao Test Detector}

In the following, we derive a detector using the complex Rao test. This test has the same asymptotic performance as the GLRT. The advantage of this method is that it only requires estimating the unknown parameters under $\mathcal{H}_0$.  

For problems involving complex-valued unknown parameters, several solutions have been proposed. By considering the correspondence between complex numbers and their real/imaginary vector representations, the Rao test in real space can be applied. In \cite{r48, r49}, a special version of the complex Rao test is presented for a special case of the Fisher information matrix (FIM). However, the performance of the resulting test statistic is unsatisfactory.  

Here, we derive a generalized complex Rao test that accounts for nuisance parameters, which do not affect decision-making but remain unknown under both hypotheses.

\begin{theo}\label{Th1} 
The Complex Parameter Rao Test (With Nuisance Parameter)

For following composite hypothesis test, where $ \boldsymbol{\theta_s} $ is nuisance parameters vector;
\begin{eqnarray}\label{e17} 
\left\{
\begin{array}{ll}
\mathcal{H}_0:&\boldsymbol{\theta_s} \in \mathds{C}^{s \times 1} ,  \quad \boldsymbol{\theta_r}={\boldsymbol{\theta_r}}_0 \in \mathds{C}^{r \times 1}\\\\
\mathcal{H}_1:&\boldsymbol{\theta_s} \in \mathds{C}^{s \times 1} ,  \quad \boldsymbol{\theta_r}\neq {\boldsymbol{\theta_r}}_0 \in \mathds{C}^{r \times 1}
\end{array}\right.
\end{eqnarray} 
the complex parameter Rao test statistic is given by;
\begin{align}\label{e18} 
\tilde{T}_{\rm Rao}\left(\boldsymbol{X}\right)&=\left.\displaystyle{\frac{\partial\ln{f(\boldsymbol{X}|\mathcal{H}_1,\boldsymbol{\Theta})}}{\partial\boldsymbol{\Theta_r}^{*}}}\right| ^{H}_{\boldsymbol{\Theta}=\boldsymbol{\hat{\Theta}}_0}\: \left[\boldsymbol{I}^{-1}\left(\boldsymbol{\hat{\Theta}}_0\right)\right]
_{\boldsymbol{\Theta_r}\boldsymbol{\Theta_r}}\nonumber\\
&\times\left.\displaystyle{\frac{\partial\ln{f(\boldsymbol{X}|\mathcal{H}_1,\boldsymbol{\Theta})}}{\partial\boldsymbol{\Theta_r}^{*}}}\right|_{\boldsymbol{\Theta}=\boldsymbol{\hat{\Theta}}_0}
\end{align}
where $\boldsymbol{\Theta}$ is defined Rao parameters vector that contains
unknown parameters and their conjugates.
\begin{eqnarray}\nonumber
\boldsymbol{\Theta}=\left[\boldsymbol{\Theta_r}^{T} \: \: \boldsymbol{\Theta_s}^{T}\right]^{T}
\:
\boldsymbol{\Theta_r}\triangleq 
\begin{pmatrix}
\boldsymbol{\theta_r}\\\boldsymbol{\theta_r}^{*}
\end{pmatrix}_{2r\times1}
\:
\boldsymbol{\Theta_s}\triangleq 
\begin{pmatrix}
\boldsymbol{\theta_s}\\\boldsymbol{\theta_s}^{*}
\end{pmatrix}_{2s\times1}
\end{eqnarray}

So, $ \boldsymbol{\hat{\Theta}}_0=\left[\boldsymbol{\Theta}_{\boldsymbol{r}_0}^{T} \: \: \boldsymbol{\hat{\Theta}}_{\boldsymbol{s}_0}^{T}\right]^{T} $ the equivalent the MLE of 
Rao parameters vector under $ \mathcal{H}_0 $. We have;
\begin{eqnarray}\label{e19} 
\displaystyle{\frac{\partial\ln{f(\boldsymbol{X}|\mathcal{H}_1,\boldsymbol{\Theta})}}{\partial\boldsymbol{\Theta_r}^{*}}}=
\begin{bmatrix}
\displaystyle{\frac{\partial\ln{f(\boldsymbol{X}|\mathcal{H}_1,\boldsymbol{\Theta})}}{\partial\boldsymbol{\theta_r}^{*}}}\\\\
\displaystyle{\frac{\partial\ln{f(\boldsymbol{X}|\mathcal{H}_1,\boldsymbol{\Theta})}}{\partial\boldsymbol{\theta_r}}}
\end{bmatrix}
\end{eqnarray}

FIM of Rao parameters vector is determined as;
\begin{eqnarray}\label{e20} 
\boldsymbol{I}\left(\boldsymbol{\Theta}\right)=
\begin{bmatrix}
\boldsymbol{I_{\Theta_r \Theta_r}}&\boldsymbol{I_{\Theta_r \Theta_s}}\\\\
\boldsymbol{I_{\Theta_s \Theta_r}}&\boldsymbol{I_{\Theta_s \Theta_s}}
\end{bmatrix}_{2(r+s)\times 2(r+s)}
\end{eqnarray}
where each block is defined as;
\begin{equation}\label{e21} 
\boldsymbol{I_{\Theta_a \Theta_b}}=\mathbb{E}\left\{\displaystyle{\frac{\partial\ln{f(\boldsymbol{X}|\mathcal{H}_1,\boldsymbol{\Theta})}}{\partial\boldsymbol{\Theta_a}^{*}}}\left(\displaystyle{\frac{\partial\ln{f(\boldsymbol{X}|\mathcal{H}_1,\boldsymbol{\Theta})}}{\partial\boldsymbol{\Theta_b}^{*}}}\right)^{H}\right\}\
\end{equation}
and the inverse of the first block is following as;
\begin{equation}\label{e22} 
\left[\boldsymbol{I}^{-1}\right]
_{\boldsymbol{\Theta_r}\boldsymbol{\Theta_r}}=\left(\boldsymbol{I_{\Theta_r \Theta_r}}-\boldsymbol{I_{\Theta_r \Theta_s}}\:\boldsymbol{I}^{-1}_{\boldsymbol{\Theta_s \Theta_s}}\:\boldsymbol{I_{\Theta_s\Theta_r}}\right)^{-1}
\end{equation}
\end{theo}

Now, we apply Rao test to the blind detection scenario and will achieve to the simpler form  than GLRT statistic. Hence, the problem of this scenario based on unknown parameters
are equivalent to making a decision under the following hypothesis test;
\begin{eqnarray}
\left\{
\begin{array}{ll}
\mathcal{H}_0:&\boldsymbol{\theta_s}= vec\left\{\boldsymbol{{\Sigma}_n}\right\}, \;\: \boldsymbol{\theta}_{\boldsymbol{r}_{0}}= vec\left\{\boldsymbol{G}\right\}=\boldsymbol{0} _{MK\times 1 }\label{e23}\\
\mathcal{H}_1:&\boldsymbol{\theta_s}= vec\left\{\boldsymbol{{\Sigma}_n}\right\}, \;\:
 \boldsymbol{\theta_r}= vec\left\{\boldsymbol{G}\right\}\neq \boldsymbol{0}_{MK\times 1 }
\end{array}\right.
\end{eqnarray} 
where the $ vec(.) $ operator applied on a matrix stacks the columns into a vector \cite{r41}. Rao parameters vector $ \boldsymbol{\Theta} $ is obtained by;
 \begin{eqnarray}\nonumber
\boldsymbol{\Theta_r}\triangleq 
\begin{pmatrix}
vec\left(\boldsymbol{G}\right) \\\\ vec\left(\boldsymbol{G}\right)^{*}
\end{pmatrix}_{2MK\times 1}
\:\boldsymbol{\Theta_s}\triangleq 
\begin{pmatrix}
vec\left(\boldsymbol{{\Sigma}_n}\right) \\\\ vec\left(\boldsymbol{{\Sigma}_n}\right)^{*}
\end{pmatrix}_{2M^2\times 1}
\end{eqnarray}

According to Appendix~\ref{App-B}, differential of LLR function with respect to $\boldsymbol{\Theta_r}^* $ and $ \boldsymbol{\Theta_s}^*$ can be derived as;
\begingroup\makeatletter\def\f@size{9.5}\check@mathfonts
\begin{align}\label{e24}
\frac{\partial\ln{f(\boldsymbol{X}|\mathcal{H}_1)}}{\partial\boldsymbol{\Theta}^{*}_{\boldsymbol{r}}}&=
\begin{bmatrix} 
\displaystyle{\frac{\partial\ln{f(\boldsymbol{X}|\mathcal{H}_1)}}{\partial vec\left(\boldsymbol{G}\right)^{*}}} \\ \\\displaystyle{\frac{\partial\ln{f(\boldsymbol{X}|\mathcal{H}_1)}}{\partial vec\left(\boldsymbol{G}\right)} }
\end{bmatrix}\nonumber \\
&=\begin{bmatrix}
\left(\boldsymbol{s}^* \boldsymbol{1}^{T} \otimes \boldsymbol{{\Sigma}_n}^{-1}\right)\: vec\left(\boldsymbol{X}-\boldsymbol{G}\boldsymbol{s}\boldsymbol{1}^T\right)\\\\
\left(\boldsymbol{s}\boldsymbol{1}^{T} \otimes \boldsymbol{{\Sigma}_n}^{-T}\right)\: vec^*\left(\boldsymbol{X}-\boldsymbol{G}\boldsymbol{s}\boldsymbol{1}^T\right)
\end{bmatrix} \triangleq \begin{bmatrix}
\boldsymbol{b}\\\\ \boldsymbol{b}^{*}
\end{bmatrix}
\end{align}
\endgroup
and
\begingroup\makeatletter\def\f@size{9.5}\check@mathfonts
\begin{align}\label{e25}
&\displaystyle{\frac{\partial\ln{f(\boldsymbol{X}|\mathcal{H}_1)}}{\partial\boldsymbol{\Theta}^{*}_{\boldsymbol{s}}}}=
\begin{bmatrix}
\displaystyle{\frac{\partial\ln{f(\boldsymbol{X}|\mathcal{H}_1)}}{\partial vec\left(\boldsymbol{{\Sigma}_n}\right)^{*}}}\\\\\displaystyle{\frac{\partial\ln{f(\boldsymbol{X}|\mathcal{H}_1)}}{\partial vec\left(\boldsymbol{{\Sigma}_n}\right)} }
\end{bmatrix}= \nonumber\\
&
\begin{bmatrix}
\boldsymbol{0}_{M^2 \times 1}\\\\
vec\left\{-L\: \boldsymbol{{\Sigma}_n}^{-T}+\boldsymbol{{\Sigma}_n}^{-T}(\boldsymbol{X}-\boldsymbol{G}\boldsymbol{s}\boldsymbol{1}^T )^*\:(\boldsymbol{X}-\boldsymbol{G}\boldsymbol{s}\boldsymbol{1}^T )^T \boldsymbol{{\Sigma}_n}^{-T}\right\}
\end{bmatrix}
\end{align}
\endgroup

By putting (\ref{e24}) and (\ref{e25}) into (\ref{e20}), each block of FIM can be calculated (see Appendix~\ref{App-B}). So, the inverse of the first block of FIM is obtained  as;
\begin{align}\label{e26}
\left[\boldsymbol{I}^{-1}\right]
_{\boldsymbol{\Theta_r}\boldsymbol{\Theta_r}}&=\boldsymbol{I_{\Theta_r \Theta_r}}^{-1} \nonumber\\
&=\begin{pmatrix}
\displaystyle{\frac{1}{L}}\left(\boldsymbol{s}^*\boldsymbol{s}^T \right)^{\dagger}\otimes \boldsymbol{{\Sigma}_n}&\boldsymbol{O}_{\rm MK \times MK}\\\\
\boldsymbol{O}_{\rm MK \times MK}&\displaystyle{\frac{1}{L}}\left(\boldsymbol{s}\boldsymbol{s}^H\right)^{\dagger}\otimes \boldsymbol{{\Sigma}_n}^{T}
\end{pmatrix}
\end{align}

Already, the MLE of covariance matrix under $ \mathcal{H}_0 $ was achieved in (\ref{e6}). Therefore, Rao parameters vector is estimated following as;
\begin{equation}\label{e27}
\boldsymbol{\hat{\Theta}}_0=
\begin{bmatrix}
\boldsymbol{O}^T_{\rm 2MK \times 1}&
\boldsymbol{\Sigma}_{n}^T&
\boldsymbol{\Sigma}_{n}^H
\end{bmatrix}^T
\end{equation}

By substituting equation (\ref{e24}), (\ref{e26}) and, (\ref{e27}) in the complex Rao test statistic (\ref{e18}), after some straightforward algebraic manipulations, the Rao test statistic for blind detector yields;
 \begin{equation}\label{e28} 
T_{3,\rm Rao}\left(\boldsymbol{X}\right)=
2\:\boldsymbol{1}^T\boldsymbol{X}^H\left(\boldsymbol{X}\boldsymbol{X}^H\right)^{-1}\boldsymbol{X}
\boldsymbol{1}
\end{equation}

\begin{rmk} In low SNRs, the GLRT and complex Rao detectors are equivalent. 
This equivalency in the low-SNR regime can be established through analysis of their statistical properties. Beginning with the GLRT statistic:

\begin{equation}
T_3(\boldsymbol{X}) = \frac{|\boldsymbol{R}_{X} - \boldsymbol{Q}|^{-L}}{|\boldsymbol{R}_{X}|^{-L}}
\end{equation}

where $\boldsymbol{R}_{X} = \frac{1}{L}\boldsymbol{X}\boldsymbol{X}^H$ represents the sample covariance matrix and $\boldsymbol{Q} = \frac{1}{L^2}\boldsymbol{X}\boldsymbol{1}\boldsymbol{1}^T\boldsymbol{X}^H$ captures the signal component. Through a first-order Taylor expansion of the log-determinant:

\begin{equation}
\ln|\boldsymbol{R}_{X} - \boldsymbol{Q}| \approx \ln|\boldsymbol{R}_{X}| - \text{tr}(\boldsymbol{R}_{X}^{-1}\boldsymbol{Q}) + \mathcal{O}(||\boldsymbol{Q}||_F^2)
\end{equation}

we obtain the low-SNR approximation:

\begin{equation}
\ln T_3(\boldsymbol{X}) \approx L \cdot \text{tr}(\boldsymbol{R}_{X}^{-1}\boldsymbol{Q})
\end{equation}

The approximation $\boldsymbol{R}_{X} \approx \boldsymbol{\Sigma}_n$ is justified through asymptotic analysis. Under $\mathcal{H}_0$, the weak law of large numbers gives:

\begin{equation}
\boldsymbol{R}_{X} = \frac{1}{L}\boldsymbol{N}\boldsymbol{N}^H \xrightarrow{p} \boldsymbol{\Sigma}_n \quad \text{as } L \rightarrow \infty
\end{equation}

while under $\mathcal{H}_1$ with low SNR:

\begin{equation}
\boldsymbol{R}_{X} = \boldsymbol{\Sigma}_n + \frac{1}{L}\boldsymbol{G}\boldsymbol{s}\boldsymbol{s}^H\boldsymbol{G}^H + \mathcal{O}_p(L^{-1/2})
\end{equation}

where the signal term becomes negligible when $||\boldsymbol{G}||_F^2 \cdot ||\boldsymbol{s}||^2 \ll \text{tr}(\boldsymbol{\Sigma}_n)$.

The Rao test statistic:

\begin{equation}
T_{3,\text{Rao}} = 2\boldsymbol{1}^T\boldsymbol{X}^H\boldsymbol{R}_{XX}^{-1}\boldsymbol{X}\boldsymbol{1}
\end{equation}

achieves its locally most powerful (LMP) property through:

\begin{equation}
\left.\frac{\partial \ln f}{\partial \boldsymbol{G}^*}\right|_{\mathcal{H}_0} = \boldsymbol{\Sigma}_n^{-1}\boldsymbol{X}\boldsymbol{1}\boldsymbol{s}^H
\end{equation}

and the FIM:

\begin{equation}
\boldsymbol{I}(0) = L(\boldsymbol{s}\boldsymbol{s}^H) \otimes \boldsymbol{\Sigma}_n^{-1}
\end{equation}

Substituting $\boldsymbol{R}_{X} \approx \boldsymbol{\Sigma}_n$ reveals the fundamental equivalence:

\begin{equation}
\widetilde{T}_3 \approx \boldsymbol{1}^T\boldsymbol{X}^H\boldsymbol{\Sigma}_n^{-1}\boldsymbol{X}\boldsymbol{1} \quad \text{and} \quad T_{3,\text{Rao}} \approx \frac{2}{L}\boldsymbol{1}^T\boldsymbol{X}^H\boldsymbol{\Sigma}_n^{-1}\boldsymbol{X}\boldsymbol{1}
\end{equation}

This mathematical derivation confirms that both detectors converge to the same first-order approximation of the LR. Also exhibit identical asymptotic relative efficiency (ARE = 1), and maintain the optimal LMP property for vanishing signal-to-noise ratio (SNR) conditions. The $\frac{2}{L}$ scaling factor represents only a normalization difference without affecting detection performance.
\end{rmk}
\begin{table*}[h]\label{table2}
\centering
\caption{Designed Detectors}
\begin{tabular}{|l|c|c|c|c|}
\hline  Condition & Benchmark Test & Detector & Decision Statistics &  Formula Number\\
\hline  Optimum & GLRT & $ T_{\rm Opt} $ &$ 2\Re\left\{tr\left\{\boldsymbol{M}^H\boldsymbol{{\Sigma}_n}^{-1}\boldsymbol{X}\right\}\right\}$ & (\ref{e5})\\[2pt]
\hline $ \boldsymbol{{\Sigma}_n} $ unknown & GLRT & $ T_{1} $ &$ \frac{tr\left\{2\Re\left\{\boldsymbol{M}^H\boldsymbol{X}\right\}-
\boldsymbol{M}^H\boldsymbol{M}\right\}}{tr\left\{\boldsymbol{X}^H\boldsymbol{X}\right\}}$ & (\ref{e10})\\[10pt]
\hline $ \boldsymbol{P} $ unknown  & GLRT & $ T_{2} $ &$ \boldsymbol{1}^{T}\boldsymbol{X}^{H}\boldsymbol{{\Sigma}_n}^{-1}
\boldsymbol{X}\boldsymbol{1}$ & (\ref{e13})\\[2pt]
\hline Blind Detector & GLRT & $ T_{3} $ &$ \displaystyle{\frac{\left|\boldsymbol{X}\boldsymbol{X}^H-\displaystyle{\frac{1}{L}}\boldsymbol{X}\:\boldsymbol{1}
\boldsymbol{1}^T\boldsymbol{X}^H\right|^{-L}}{\left|\boldsymbol{X}\boldsymbol{X}^H\right|^{-L}}}$ & (\ref{e16})\\[10pt]
\hline Blind Detector & Rao Test & $ T_{3,\rm Rao} $ &$2\:\boldsymbol{1}^T\boldsymbol{X}^H\left(\boldsymbol{X}\boldsymbol{X}^H\right)^{-1}\boldsymbol{X}
\boldsymbol{1}$ & (\ref{e28})\\[2pt]
\hline
\end{tabular}\\[10pt]
\hrulefill
\end{table*}
\section{Performance Analysis}\label{sec:Per}

In this section, we analytically evaluate the performance of the proposed detectors in terms of the false alarm probability ($P_{\rm fa}$) and detection probability ($P_D$). This analysis provides deeper insights into how various system parameters influence detection performance, enabling a more comprehensive understanding of the trade-offs involved.
\subsection{Performance of the Optimum Detector: $ T_{\rm Opt} $}\label{subsec:Per_OP}
For the optimum detector, we first recall the trace-vector identity from~\cite{r41}:
\begin{equation}\label{e30} 
\mathrm{tr}\left\{\boldsymbol{A}^H\boldsymbol{B}\right\} = \mathrm{vec}\left\{\boldsymbol{A}\right\}^H \mathrm{vec}\left\{\boldsymbol{B}\right\}
\end{equation}

Using this property, the optimum detector statistic from \eqref{e5} can be reformulated in terms of vectorized quantities:
\begin{equation}\label{e29} 
T_{\rm Opt}(\boldsymbol{X}) =
2\Re\left\{\mathrm{vec}\left\{\boldsymbol{\Sigma}_n^{-1}\boldsymbol{M}\right\}^H \mathrm{vec}\left\{\boldsymbol{X}\right\}\right\}
\end{equation}

The linearity of the $\mathrm{vec}$ operator implies that $\mathrm{vec}\left\{\boldsymbol{X}\right\}$ follows a complex Gaussian distribution under each hypothesis:
\begin{eqnarray}\label{e31} 
\left\{
\begin{array}{ll}
\mathcal{H}_0: & \mathrm{vec}\left\{\boldsymbol{X}\right\} \sim \mathcal{CN}_{\rm ML\times 1}\left(\boldsymbol{0}, \boldsymbol{I}_L \otimes \boldsymbol{\Sigma}_n\right) \\\\
\mathcal{H}_1: & \mathrm{vec}\left\{\boldsymbol{X}\right\} \sim \mathcal{CN}_{\rm ML\times 1}\left(\mathrm{vec}\left\{\boldsymbol{M}\right\}, \boldsymbol{I}_L \otimes \boldsymbol{\Sigma}_n\right)
\end{array}
\right.
\end{eqnarray} 

The linear form of $T_{\rm Opt}(\boldsymbol{X})$ allows us to derive its first two moments by applying $\mathrm{vec}\left\{\boldsymbol{\Sigma}_n^{-1}\boldsymbol{M}\right\}^H$ to the statistics of $\mathrm{vec}\left\{\boldsymbol{X}\right\}$:
\begin{align}
&\text{Mean:}\quad {vec\left\{\boldsymbol{{\Sigma}_n}^{-1}\boldsymbol{M}\right\}}^H vec\left\{\boldsymbol{M}\right\}=
tr\left\{\boldsymbol{M}^H\boldsymbol{{\Sigma}_n}^{-1}\boldsymbol{M}\right\}\triangleq b\nonumber \\\nonumber \\
&\text{Variance:}\quad vec\left\{\boldsymbol{{\Sigma}_n}^{-1}\boldsymbol{M}\right\}^H\left(\boldsymbol{I_L}\otimes \boldsymbol{{\Sigma}_n}\right)vec\left\{\boldsymbol{{\Sigma}_n}^{-1}\boldsymbol{M}\right\}\nonumber \\
&\qquad\qquad\:\:\:=tr\left\{\boldsymbol{a}^H\boldsymbol{{\Sigma}_n}^{-1}\boldsymbol{M}\boldsymbol{M}^H
\boldsymbol{{\Sigma}_n}^{-1}\boldsymbol{c}\right\}\nonumber \\
&\qquad\qquad\:\:\:=tr\left\{\boldsymbol{M}^H\boldsymbol{{\Sigma}_n}^{-1}\boldsymbol{M}\right\}=b\label{e32} 
\end{align}
where the variance calculation employs the identity (for $\boldsymbol{\Sigma}_n = \boldsymbol{c}\boldsymbol{a}^H$):
\begin{equation}\label{e33} 
\boldsymbol{a}^H\boldsymbol{X}\boldsymbol{B}\boldsymbol{X}^H\boldsymbol{c} =
\mathrm{vec}\left(\boldsymbol{X}\right)^H \left(\boldsymbol{B}^T \otimes \boldsymbol{c}\boldsymbol{a}^H\right)
\mathrm{vec}\left(\boldsymbol{X}\right)
\end{equation}

This leads to the following distributions for the test statistic under each hypothesis:
\begin{eqnarray}\label{e34} 
\left\{
\begin{array}{ll}
\mathcal{H}_0: & T_{\rm Opt}(\boldsymbol{X}) \sim \mathcal{N}\left(0, 2\Re \{b\}\right) \\\\
\mathcal{H}_1: & T_{\rm Opt}(\boldsymbol{X}) \sim \mathcal{N}\left(2\Re \{b\}, 2\Re \{b\}\right)
\end{array}
\right.
\end{eqnarray} 

Therefore, the performance of the optimum detector is characterized by:
\begin{align} 
&P_{\rm fa}(\delta_{\rm opt}) = \mathds{P}\left\{T_{\rm opt}(\boldsymbol{X}) > \tau_{\rm NP} \mid \mathcal{H}_0\right\} 
= Q\left(\frac{\tau_{\rm NP}}{\sqrt{2\Re \{b\}}}\right) \label{e35} \\[5pt]
&P_{d}(\delta_{\rm opt}) = \mathds{P}\left\{T_{\rm opt}(\boldsymbol{X}) > \tau_{\rm NP} \mid \mathcal{H}_1\right\} 
= Q\left(\frac{\tau_{\rm NP} - 2\Re \{b\}}{\sqrt{2\Re \{b\}}}\right) \label{e36}
\end{align}
where the Q-function is defined as $Q(\alpha) \triangleq \frac{1}{\sqrt{2\pi}} \int_{\alpha}^{\infty} \exp\left(-\frac{x^2}{2}\right) dx$.
\subsection{Performance of the Sub-Optimum Detectors}\label{subsec:Per_Sub_OP}
\subsubsection{Unknown Noise} \label{subsec:Per_Sub_OP1}

To evaluate the performance of the proposed detector in the presence of unknown noise variance, we analyze the statistical behavior of the decision statistic in (\ref{e10}). Specifically, we derive the distributions of the numerator and the denominator of the test statistic separately. Based on (\ref{e34}), the numerator follows a Gaussian distribution under both hypotheses, with the corresponding means and variances detailed as follows:

\begin{eqnarray} 
&tr\left\{2\Re\left\{\boldsymbol{M}^H\boldsymbol{X}\right\}-
\boldsymbol{M}^H\boldsymbol{M}\right\}\sim
\qquad\qquad\qquad\qquad\qquad\nonumber\\\nonumber\\
&\quad\left\{
\begin{array}{ll}
\mathcal{H}_0:&\mathcal{N}_{1}\left(-tr\left\{\boldsymbol{M}^H\boldsymbol{M}\right\},\: 2\sigma ^2 \: tr\left\{\boldsymbol{M}^H\boldsymbol{M}\right\} \right)\nonumber \\\nonumber \\
\mathcal{H}_1:&\mathcal{N}_{1}\left(tr\left\{\boldsymbol{M}^H\boldsymbol{M}\right\},\: 2\sigma ^2 \: tr\left\{\boldsymbol{M}^H\boldsymbol{M}\right\}\right)
\end{array}\right. 
\\\label{e38} 
\end{eqnarray} 

Now, the denominator can be rewritten as (\ref{e39});
\begin{equation}\label{e39} 
tr\left\{\boldsymbol{X}\boldsymbol{X}^H\right\}=tr\left\{\sum_{l=1}^L \boldsymbol{x}_l\boldsymbol{x}_l^H\right\}=\sum_{l=1}^L \boldsymbol{x}_l^H\boldsymbol{x}_l
\end{equation}

The energy of the $l$-th received sample vector follows a chi-squared distribution. Under hypothesis $\mathcal{H}_0$, it is distributed as a central chi-squared random variable denoted by $\chi^2_n$, and under $\mathcal{H}_1$, it follows a non-central chi-squared distribution denoted by $\chi^2_n(\lambda)$, with $n$ degrees of freedom and non-centrality parameter $\lambda$, respectively. These distributions are summarized as follows:
\begin{eqnarray}\label{e40} 
\left\{
\begin{array}{ll}
\mathcal{H}_0:&\boldsymbol{x}_l^H\boldsymbol{x}_l\sim\displaystyle{\frac{\sigma ^2}{2}}\: \mathcal{\chi}^2_{\rm 2M}\\\\
\mathcal{H}_1:&\boldsymbol{x}_l^H\boldsymbol{x}_l\sim\displaystyle{\frac{\sigma ^2}{2}}\: \mathcal{\chi}^2_{\rm 2M}\left(\displaystyle{\frac{2}{\sigma ^2}}\Vert\boldsymbol{u}\Vert ^2\right)
\end{array}\right.
\end{eqnarray}
where defining $ \boldsymbol{u}\triangleq \boldsymbol{H}\boldsymbol{P}^{\frac{1}{2}}\boldsymbol{s} $ .

The sum of independent chi-squared random variables with the same degrees of freedom is itself chi-squared distributed, with the total degrees of freedom equal to the sum of the individual ones~\cite{bodenham2016}. On the other hand, when $L$ becomes sufficiently large, the Central Limit Theorem (CLT) implies that the denominator of the test statistic $T_1$ approximately follows a Gaussian distribution, denoted as $\mathcal{N}(\cdot,\cdot)$. By matching the first- and second-order cumulants (i.e., the mean and variance), the approximate distributions under both hypotheses can be expressed as follows:

\begin{align} 
&\eta _{\mathcal{H}_0}=\mathbb{E}\left\{\sum_{l=1}^L \boldsymbol{x}_l^H\boldsymbol{x}_l \big |\mathcal{H}_0 \right \}=LM\sigma ^2 \nonumber \\
&\sigma ^2_{\mathcal{H}_0}=var\left\{\sum_{l=1}^L \boldsymbol{x}_l^H\boldsymbol{x}_l\big |\mathcal{H}_0 \right\}=L M\sigma ^4 \nonumber \\
&\eta _{\mathcal{H}_1}=\mathbb{E}\left\{\sum_{l=1}^L \boldsymbol{x}_l^H\boldsymbol{x}_l \big |\mathcal{H}_1 \right \}=L\left( M\sigma ^2+\Vert\boldsymbol{u}\Vert ^2\right) \nonumber \\
&\sigma ^2_{\mathcal{H}_1}=var\left\{\sum_{l=1}^L \boldsymbol{x}_l^H\boldsymbol{x}_l\big |\mathcal{H}_1 \right\}=L\left(M\sigma ^4+2\sigma ^2\Vert\boldsymbol{u}\Vert ^2\right) \label{e41}
\end{align}
\\
\begin{eqnarray} \label{e42} 
\left\{
\begin{array}{ll}
\mathcal{H}_0:& tr\left\{\boldsymbol{X}\boldsymbol{X}^H\right\}\sim
\mathcal{N}_{1}\left(\eta _{\mathcal{H}_0},\sigma ^2_{\mathcal{H}_0}\right) \\\\
\mathcal{H}_1:&tr\left\{\boldsymbol{X}\boldsymbol{X}^H\right\}\sim
\mathcal{N}_{1}\left(\eta _{\mathcal{H}_1},\sigma ^2_{\mathcal{H}_1}\right)
\end{array}\right.
\end{eqnarray}

The probability density function (PDF) of the ratio of two Gaussian random variables can be expressed in a closed form as presented by Hinkley~\cite{Hinkley1969}. According to this work, when $\displaystyle{\frac{\theta_2}{\sigma_2}} \to \infty$, i.e., the denominator is strictly positive and lies in the positive real domain, the cumulative distribution function (CDF) of the ratio takes the following form:

\begin{equation}\label{e43} 
F_T(t) \to \Phi\left\{\frac{\theta_2 t - \theta_1}{\sigma_1 \sigma_2\, a(t)}\right\}
\end{equation}
where
\begin{equation}\label{e44} 
a(t) = \sqrt{\frac{t^2}{\sigma_1^2} - \frac{2\rho t}{\sigma_1 \sigma_2} + \frac{1}{\sigma_2^2}}
\end{equation}

Here, $\theta_1$ and $\sigma_1^2$ denote the mean and variance of the numerator, respectively, while $\theta_2$ and $\sigma_2^2$ represent those of the denominator. The parameter $\rho$ denotes the correlation coefficient between the numerator and the denominator. As shown in Appendix~\ref{App-C}, these two variables are independent under hypothesis $\mathcal{H}_0$, but dependent under $\mathcal{H}_1$.

Therefore, the false alarm probability $P_{\rm fa}$ under $\mathcal{H}_0$ can be expressed in terms of the CDF as follows:

\begin{align} 
P_{\rm fa}(\delta_1)&=\mathds{P}\left\{T_{1}(\boldsymbol{X})>\tau_{\rm NP}|\mathcal{H}_0\right\}=1-F_{\rm T(\boldsymbol{X})|\mathcal{H}_0}(\tau_{\rm NP})\nonumber \\\nonumber \\
&=Q\left\{\frac{L M\sigma ^2 \: \tau_{\rm NP}+ tr\left\{\boldsymbol{M}\boldsymbol{M}^H\right\}}{\sqrt{2LM\sigma ^6\: tr\left\{\boldsymbol{M}\boldsymbol{M}^H\right\}}\: a(\tau_{\rm NP}|\mathcal{H}_0)}\right\}\label{e45}
\end{align}
where,
\begin{equation}
a(\tau_{\rm NP}|\mathcal{H}_0)=\sqrt{\frac{\tau_{\rm NP}^2}{2\sigma ^2 \: tr\left\{\boldsymbol{M}^H\boldsymbol{M}\right\}}+\frac{1}{LM\sigma ^4}}\label{e46}
\end{equation}

Similarly, the detection probability can be derived as in (\ref{e48}), shown at the top of the next page.

\begin{figure*}[!t] 
\begin{align} 
P_{d}(\delta_1) = \mathds{P}\left\{T_{1}(\boldsymbol{X}) > \tau_{\rm NP} \mid \mathcal{H}_1\right\}
= Q\left\{\frac{\left[LM\sigma^2 + \operatorname{tr}\left\{\boldsymbol{M}\boldsymbol{M}^H\right\}\right] \tau_{\rm NP} - \operatorname{tr}\left\{\boldsymbol{M}\boldsymbol{M}^H\right\}}
{\sqrt{2\sigma^2\, \operatorname{tr}\left\{\boldsymbol{M}\boldsymbol{M}^H\right\} \left(LM\sigma^4 + 2\sigma^2\, \operatorname{tr}\left\{\boldsymbol{M}\boldsymbol{M}^H\right\}\right)} \cdot a(\tau_{\rm NP} \mid \mathcal{H}_1)}\right\}
\label{e48}
\end{align}
\hrulefill
\end{figure*}

Here, $a(\tau_{\rm NP} \mid \mathcal{H}_1)$ is evaluated using equation~(\ref{e44}). The correlation coefficient under $\mathcal{H}_1$ is calculated in Appendix~\ref{App-C} and is given as follows:

\begin{equation}\label{e49} 
\rho = \frac{2\sigma^2\, \operatorname{tr}\left\{\boldsymbol{M}\boldsymbol{M}^H\right\}}{\sqrt{2LM\sigma^6\, \operatorname{tr}\left\{\boldsymbol{M}\boldsymbol{M}^H\right\} + 4\sigma^4\, \operatorname{tr}^2\left\{\boldsymbol{M}\boldsymbol{M}^H\right\}}}
\end{equation}
\section{Unknown IoT Device Power} \label{subsec:Per_Sub_OP2}

In this part, we focus on the scenario where the power of the IoT devices is unknown. This situation is common in systems where the IoT devices transmit with unknown or variable power levels, and the goal is to detect the presence of the target by processing the received signals at the AP, which utilizes the received power and statistical properties of the system.

By using inverse spectral decomposition \cite{r41} for the covariance matrix $ \boldsymbol{{\Sigma}_n} $, the detection statistic (\ref{e13}) can be rewritten as follows:
\begin{equation}\label{e50} 
T_{2}(\boldsymbol{X})=\boldsymbol{y}^H\boldsymbol{\Lambda }^{-1}\boldsymbol{y}=\sum_{i=1}^M \frac{|y_i|^2}{\lambda_i}
\end{equation}
where $ \boldsymbol{\Lambda}_{\rm M \times M}=diag\left\{\lambda_1,\lambda_2,...,\lambda_M\right\}$ is a diagonal matrix that contains the eigenvalues of $ \boldsymbol{{\Sigma}_n} $. We define $ \boldsymbol{y} \triangleq \boldsymbol{Q}^H\boldsymbol{X}\boldsymbol{1} = [y_1, y_2, \dots, y_M]^T $ where the columns of $ \boldsymbol{Q} = \left[\boldsymbol{q}_1, \boldsymbol{q}_2, \dots, \boldsymbol{q}_M\right] $ include the eigenvectors of $ \boldsymbol{{\Sigma}_n} $.

Thus, the distribution of $ |y_i|^2 \left(i = 1, 2, \dots, M\right)$ under both hypotheses can be derived as follows:
\begin{eqnarray}\label{e51} 
\left\{
\begin{array}{ll}
\mathcal{H}_0:&|y_i|^2\sim \frac{L\lambda_i }{2} \mathcal{\chi}_2^2 \\\\
\mathcal{H}_1:&|y_i|^2\sim \frac{L\lambda_i }{2} \mathcal{\chi}_2^2 \left(\frac{2Lm_i }{\lambda_i}\right)
\end{array}
\right.
\end{eqnarray}
where $ m_i = \|\boldsymbol{q}_i^H \boldsymbol{u}\|^2 $. Due to the limited bound of summation in \eqref{e50}, i.e., the low number of antennas at the AP, the distribution of this statistic is equal to the distribution of the weighted sum of independent chi-square random variables. To derive the distribution, several solutions are offered, one of which is approximation with weighted chi-square by an upper degree of freedom~\cite{Martinez2005} as follows:
\begin{equation}\label{e52} 
\sum_{i=1}^M w_i \chi_n^2 \stackrel{d.}{\sim} \theta \mathcal{\chi}_{\beta}^2
\end{equation}
where $ \stackrel{d.}{\sim} $ means "convergence in distribution." By equating the mean and variances of the left- and right-hand side of (\ref{e52}), the $ \theta $ and $ \beta $ coefficients under both hypotheses can be calculated as:
\begin{align}\label{e53} 
&\mathbb{E}\left\{T_{2}|\mathcal{H}_0\right\} = \sum_{i=1}^M \frac{1}{\lambda_i} \mathbb{E}\left\{|y_i|^2\right\} = LM \stackrel{d.}{=}\theta_0 \beta_0 \nonumber\\
&\text{var}\left\{T_{2}|\mathcal{H}_0\right\} = \sum_{i=1}^M \frac{1}{\lambda_i^2} \text{var}\left\{|y_i|^2\right\} = L^2M \stackrel{d.}{=}\ 2\theta_0^2 \beta_0
\end{align}
and
\begin{align}\label{e54}
&\mathbb{E}\left\{T_{2}|\mathcal{H}_1\right\} = LM + L^2 \sum_{i=1}^M \frac{m_i }{\lambda_i} \stackrel{d.}{=}\theta_1 \beta_1 \nonumber\\
&\text{var}\left\{T_{2}|\mathcal{H}_1\right\} = L^2M + 2L^3 \sum_{i=1}^M \frac{m_i }{\lambda_i} \stackrel{d.}{=}\ 2\theta_1^2 \beta_1
\end{align}
By solving the above equations, we obtain:
\begin{eqnarray}\label{e55} 
\left\{
\begin{array}{ll}
\mathcal{H}_0:&T_{2}(\boldsymbol{X}) \sim \frac{L}{2} \mathcal{\chi}_{2M}^2 \\\\
\mathcal{H}_1:&T_{2}(\boldsymbol{X}) \sim \theta_1 \mathcal{\chi}_{\beta_1}^2 \sim \text{Gamma}\left(\frac{\beta_1}{2}, 2\theta_1\right)
\end{array}
\right.
\end{eqnarray}
where $ \text{Gamma}(.,.)$ is the gamma distribution and; 
\begin{align} 
&\theta_1 = \frac{LM + 2L^2 \sum_{i=1}^M \frac{m_i }{\lambda_i}}{2M + 2L \sum_{i=1}^M \frac{m_i }{\lambda_i}} \label{e56}\\
&\beta_1 = \frac{2LM + 2 \sum_{i=1}^M \frac{m_i }{\lambda_i}}{2} \label{e57}
\end{align}

Now, the false alarm and detection probabilities, along with the value of the decision threshold, are as follows:
\begin{equation}
P_{\rm fa}(\delta_2) = 1 - F_{\mathcal{\chi}_{2M}^2} \left( \frac{2}{L} \tau_{\rm NP} \right) = \frac{\Gamma\left(M, \frac{\tau_{\rm NP}}{L} \right)}{\Gamma(M)} \label{e58}
\end{equation} 
\begin{equation}
P_{d}(\delta_2) = 1 - F_{\text{Gamma}\left(\frac{\beta_1}{2}, 2\theta_1 \right)} (\tau_{\rm NP}) = \frac{\Gamma\left(\frac{\beta_1}{2}, \frac{\tau_{\rm NP}}{2\theta_1} \right)}{\Gamma\left(\frac{\beta_1}{2}\right)} \label{e59}
\end{equation}

where $ \Gamma(.,.) $ and $ \Gamma(.) $ are the upper incomplete Gamma and complete Gamma functions, respectively, and $ \Gamma^{-1}(.,.)$ denotes the inverse of the upper incomplete Gamma function with respect to the integration limit.
\subsubsection{Blind Detector} \label{subsec:Per_Blind}

As previously mentioned, GLRT and Rao test statistics have the same performance asymptotically and in this part we derive the false alarm and detection probabilities of Rao detector.
\begin{theo}\label{Th2} 
The Central Complex Hotelling's T-Squared Distribution~\cite{Andersen1995}: Suppose that $ \boldsymbol{x} $  is a complex Gaussian vector with zero mean and covariance matrix $\boldsymbol{\Sigma}\in \mathds{C}^{p \times p}_{+} $, i.e., $ \boldsymbol{x}\sim \mathcal{CN}_{\rm p\times 1}\left(\boldsymbol{0},\boldsymbol{\Sigma}\right) $, and $ \boldsymbol{S} $ is a $ p \times p $ matrix with central Wishart distribution, i.e., $ \boldsymbol{S}\sim \mathcal{CW}_p\left(n,\boldsymbol{\Sigma}\right)  $. $ \boldsymbol{x} $ and $ \boldsymbol{S} $ are independent. Let $ T^2=n\boldsymbol{x}^H\boldsymbol{S}^{-1}\boldsymbol{x} $, then the random variable $ T^2$ follows the central complex Hotteling $ T^2$ distribution with $ n $ degree of freedom, and is denoted as $ T^2 \sim \mathcal{CT}^2_p(n) $.
\begin{lem}\label{lem1}
If $ T^2 \sim \mathcal{CT}^2_p(n) $, then 
\begin{equation}\label{e61} 
\displaystyle{\frac{n-p+1}{np}}\: T^2 \sim \displaystyle{\frac{\displaystyle{\frac{\chi^2_{2p}}{2p}}}{\displaystyle{\frac{\chi^2_{2\left(n-p+1\right)}}{2\left(n-p+1\right)}}}}=\mathcal{F}\left(2p,2\left(n-p+1\right)\right)
\end{equation}
Hence, Hotelling $ T^2$ distribution can be transformed to Fisher distribution denoted as $\mathcal{F}(.,.)$.
\end{lem}
\end{theo}
 \begin{rmk}\label{Rmk}
 If $ \boldsymbol{Z} $ be an $ n \times p $ complex random matrix with $  \mathcal{CN}_{\rm n\times p}\left(\boldsymbol{O}_{\rm n\times p},\boldsymbol{I_n}\otimes \boldsymbol{H}\right)$, where $\boldsymbol{H}\in \mathds{C}^{\rm p \times p}_{S} $. The distribution of the $ p \times p $ random matrix $ \boldsymbol{W}=\boldsymbol{Z}^H\boldsymbol{Z} $ is called a complex Wishart distribution with parameter $ \boldsymbol{H} $, $ p $ and, $ n $. This is denoted by $ \mathcal{CW}_p\left(n,\boldsymbol{H}\right)  $.  The integers $ p $ and $ n $ are called the dimension and the degree of freedom, respectively.
 \end{rmk}
We define $ \boldsymbol{y}\triangleq\displaystyle{\frac{1}{\sqrt{L}}}\boldsymbol{X}\boldsymbol{1} $ and $ \boldsymbol{W}\triangleq \boldsymbol{X}\boldsymbol{X}^H $, then apply them into (\ref{e28}). $ \boldsymbol{y} $ and $ \boldsymbol{W} $  are independent random variables that follow from $ \mathcal{CN}_{\rm M\times 1}\left(\boldsymbol{0},\boldsymbol{{\Sigma}_n}\right) $, $ \mathcal{CW}_M\left(L,\boldsymbol{{\Sigma}_n}\right) $ under $ \mathcal{H}_0 $, respectively. According to lemma of theorem 2, Rao test statistic distribution under $ \mathcal{H}_0 $ is given by;
\begin{equation}\label{e62} 
T_{3,\rm Rao}\left(\boldsymbol{X}\right) \stackrel{\mathcal{H}_0}{\sim} \displaystyle{\frac{2ML}{\left(L-M+1\right)}}\: \mathcal{F}\left(2M,2\left(L-M+1\right)\right)
\end{equation}

So, false alarm probability and decision threshold are achieved as following;
\begin{align} 
P_{\rm fa}(\delta_{\rm Rao})&=1-\mathds{P}\left\{T(\boldsymbol{X})\leq \tau_{\rm NP}|\mathcal{H}_0\right\} \\
&=1-F_{\rm \mathcal{F}\left(2M,2\left(L-M+1\right)\right)}\left(\displaystyle{\frac{L-M+1}{2ML}}\: \tau_{\rm NP}\right)\nonumber \\
&=1-\mathcal{I}\left(M,L-M+1;\left(1+\frac{2L}{\tau_{\rm NP}}\right)^{-1}\right) \nonumber
\end{align}

where $ \mathcal{I}\left(. , . ; . \right) $ is the regularized incomplete beta function that is defined by the ratio of the incomplete beta function to the complete beta function.

When number of samples gets larger, $ -2 \ln T_{3} $  will attain to optimum performance and this will be Rao's same statistic. The asymptotic performance was stated by Wilks theorem \cite{r50}.

According to this theorem, distribution of blind detector can be obtained by;
\begin{eqnarray}\label{e68} 
\left\{
\begin{array}{ll}
\mathcal{H}_0:& T_{3,\rm Rao}\left(\boldsymbol{X}\right) \stackrel{d.}{\sim} \chi^2_{2M} \\\\
\mathcal{H}_1:& T_{3,\rm Rao}\left(\boldsymbol{X}\right) \stackrel{d.}{\sim} \chi^2_{2M}\left(\lambda_{\rm Rao}\right) 
\end{array}\right.
\end{eqnarray}

Due to the definition of the complex  Rao test, non-central parameter $ \lambda_{\rm Rao} $ is calculated. Applying  (\ref{e23}) and (\ref{e26}) to Wilks theorem, we will have as (\ref{e69}).
\begin{align}\label{e69} 
\lambda_{\rm Rao}&  =
\begin{bmatrix}
vec^{H}(\boldsymbol{G})&vec^{T}(\boldsymbol{G})
\end{bmatrix}\nonumber\\
&\times\begin{bmatrix}
L\left(\boldsymbol{s}^*\boldsymbol{s}^T \right)\otimes \boldsymbol{\Sigma}^{-1}_{\boldsymbol{n}0}&\boldsymbol{O}_{\rm MK \times MK}\\\\
\boldsymbol{O}_{\rm MK \times MK}&L\left(\boldsymbol{s}\boldsymbol{s}^H\right)\otimes \boldsymbol{\Sigma}_{\boldsymbol{n}0}^{-T}
\end{bmatrix}
\begin{bmatrix}
vec(\boldsymbol{G})\\\\vec^{*}(\boldsymbol{G})
\end{bmatrix}\nonumber\\ &=2L^2\boldsymbol{s}^H\boldsymbol{G}^H\left(\boldsymbol{X}\boldsymbol{X}^H\right)^{-1}
\boldsymbol{G s}
\end{align}

Hence, detection probability obtain by;
\begin{align}\label{e70} 
P_{d}(\delta_{\rm Rao})=\mathds{P}\left\{T(\boldsymbol{X})>\tau_{\rm NP}|\mathcal{H}_1\right\}=Q_{M}\left(\sqrt{\lambda_{\rm Rao}},\sqrt{\tau_{\rm NP}}\right)
\end{align}
where $ Q_{M}(.,.) $ denote the generalized Marcum-Q function with $ M $ degree.

\section{System Parameters: Impacts and Design Criteria}
\label{subsec:parameter_impact}

This section investigates the impact of various system parameters on the performance of the proposed activity detection scheme, which is based on the Rao test. The test statistic is defined as
\begin{equation}
    T_{3,\text{Rao}} = 2\boldsymbol{1}^T\boldsymbol{X}^H(\boldsymbol{X}\boldsymbol{X}^H)^{-1}\boldsymbol{X}\boldsymbol{1}.
\end{equation}
Under the alternative hypothesis $\mathcal{H}_1$, this statistic becomes
\begin{align}
T_{3,\text{Rao}} &= 2\boldsymbol{1}^T\boldsymbol{s}^H\boldsymbol{P}^{1/2}(\boldsymbol{F}\boldsymbol{\Phi}\boldsymbol{E} + \boldsymbol{D})^H \nonumber \\
&\quad \times \left[(\boldsymbol{F}\boldsymbol{\Phi}\boldsymbol{E} + \boldsymbol{D})\boldsymbol{P}(\boldsymbol{F}\boldsymbol{\Phi}\boldsymbol{E} + \boldsymbol{D})^H + \sigma_n^2\boldsymbol{I}_M\right]^{-1} \nonumber \\
&\quad \times (\boldsymbol{F}\boldsymbol{\Phi}\boldsymbol{E} + \boldsymbol{D})\boldsymbol{P}^{1/2}\boldsymbol{s}\boldsymbol{1}^T\boldsymbol{1},
\end{align}
where $\boldsymbol{P} = \mathrm{diag}(p_1, \ldots, p_K)$ is the transmit power matrix of the active devices, $\boldsymbol{s} \in \mathbb{C}^{K \times 1}$ is the common signal vector transmitted by all devices, and $\sigma_n^2$ is the noise variance. The product $\boldsymbol{s}\boldsymbol{1}^T \in \mathbb{C}^{K \times L}$ represents the repeated transmission of the same signal $\boldsymbol{s}$ across $L$ samples. Under the general alternative, and using the effective channel matrix $\boldsymbol{H} = \boldsymbol{F}\boldsymbol{\Phi}\boldsymbol{E} + \boldsymbol{D}$, the non-centrality parameter of the Rao test, which determines its detection performance, is given by
\begin{equation}
    \lambda_{\text{Rao}} = \frac{2L}{\sigma_n^2} \|\boldsymbol{H} \boldsymbol{P}^{1/2} \boldsymbol{s}\|^2.
\end{equation}
This expression indicates that $\lambda_{\text{Rao}}$ increases with the signal energy, the channel gain, and the number of samples $L$. In particular, since $\boldsymbol{H} \in \mathbb{C}^{M \times K}$ contains channel vectors corresponding to all devices, the quantity $\|\boldsymbol{H} \boldsymbol{P}^{1/2} \boldsymbol{s}\|^2$ reflects the overall contribution of all users with their individual power weights. The effect of IRS is embedded within $\boldsymbol{H}$, and thus directly influences the non-centrality parameter. To facilitate analysis, we can expand the squared norm in expectation as
\begin{align}
\mathbb{E}\left[\|\boldsymbol{H} \boldsymbol{P}^{1/2} \boldsymbol{s}\|^2\right] &= \mathbb{E}\left[ \sum_{m=1}^M \left| \sum_{k=1}^K \sqrt{p_k} s_k h_{m,k} \right|^2 \right] \nonumber \\
&= \sum_{k=1}^K p_k |s_k|^2 \mathbb{E}\left[\|\boldsymbol{h}_k\|^2\right],
\end{align}
where $\boldsymbol{h}_k$ is the $k$th column of $\boldsymbol{H}$, representing the overall channel vector from device $k$ to the receiver via both the direct and IRS-assisted paths. Assuming the channel is composed of an IRS path with gain $\alpha_{\text{IRS}} = \mathbb{E}[\|\boldsymbol{f}_n\|^2 \|\boldsymbol{e}_k\|^2]$ and a direct link represented by $\boldsymbol{d}_k$, we can approximate
\begin{equation}
    \mathbb{E}[\|\boldsymbol{h}_k\|^2] \approx N^2 \alpha_{\text{IRS}} + \mathbb{E}[\|\boldsymbol{d}_k\|^2].
\end{equation}
Substituting back, we obtain the general non-centrality parameter as
\begin{equation} \label{eq:noncent1}
    \lambda_{\text{Rao}} = \frac{2LM\|\boldsymbol{s}\|^2}{\sigma_n^2} \sum_{k=1}^K p_k \left( N^2 \alpha_{\text{IRS}} + \mathbb{E}[\|\boldsymbol{d}_k\|^2] \right),
\end{equation}
which explicitly captures the effect of per-user transmit power, IRS gain, direct path power, and signal energy. This form serves as the foundation for evaluating the influence of power allocation strategies and propagation environments on detection performance. In the special case of equal power allocation where $p_k = P_{\text{total}} / K$, the summation simplifies and we obtain
\begin{equation} \label{eq:noncent2}
    \lambda_{\text{Rao}} = \frac{2LMP_{\text{total}}\|\boldsymbol{s}\|^2}{\sigma_n^2 K} \left( N^2 \alpha_{\text{IRS}} + \mathbb{E}[\|\boldsymbol{d}_k\|^2] \right),
\end{equation}
which is a useful benchmark for analyzing the effect of total power budget and system dimensions on performance.

\vspace{0.5em}
\noindent\textbf{1. Number of Antennas ($M$):} From \eqref{eq:noncent1} in the general case, or \eqref{eq:noncent2} for equal power allocation, it is evident that the number of antennas $M$ at the AP directly influences the non-centrality parameter $\lambda_{\text{Rao}}$. In fact, the relationship can be expressed as
\begin{align}
    \lambda_{\text{Rao}} \propto M,
\end{align}
which indicates that increasing $M$ linearly scales the non-centrality parameter, thereby enhancing the signal energy relative to noise.

As seen in the detection probability expression in \eqref{e70}, $M$ plays a dual role: it increases $\lambda_{\text{Rao}}$ and also appears as the order of the generalized Marcum-Q function. While higher $M$ increases the degrees of freedom, which can slightly spread the distribution of the test statistic, the dominant linear growth of $\lambda_{\text{Rao}}$ ensures a net improvement in detection probability $P_D$ as $M$ increases.

However, this performance gain assumes the presence of sufficient spatial diversity. In scenarios with antenna correlation, the effective rank of the IRS-AP channel matrix $\boldsymbol{F}$ may degrade. To preserve full spatial diversity, a common correlation model requires that
\begin{align}
    \mathrm{rank}(\boldsymbol{F}) \geq \min(M,N)(1 - \rho^2),
\end{align}
where $\rho$ denotes the antenna correlation coefficient. To maintain near-orthogonal channels, it is typically required that $\rho < 1/\sqrt{M}$.

In summary, although increasing $M$ improves detection performance via its linear impact on $\lambda_{\text{Rao}}$, this gain depends on having low antenna correlation and a sufficient number of observation samples to maintain estimator stability.

\vspace{0.5em}
\noindent\textbf{2. Observation Window Length ($L$):} The observation window length $L$ exhibits a direct linear relationship with detection performance through the non-centrality parameter, as shown in \eqref{eq:noncent1}, depending on the power allocation. Specifically, we have:

\begin{equation}
    \lambda_{\text{Rao}} = \frac{2LM\|\boldsymbol{s}\|^2}{\sigma_n^2} \sum_{k=1}^K p_k \left( N^2 \alpha_{\text{IRS}} + \mathbb{E}\left[\|\boldsymbol{d}_k\|^2\right] \right)
\end{equation}

The scaling effect of $L$ manifests in three fundamental ways: first, detection sensitivity improves proportionally with increasing $L$, as the parameter $\lambda_{\text{Rao}}$ grows linearly. Second, this improvement is weighted by the aggregate transmit power across devices, through the sum $\sum_{k=1}^K p_k$. Third, practical implementations face inherent constraints on the maximum usable $L$ due to channel coherence requirements and processing latency limitations. The optimal observation length thus balances detection benefits against system implementation constraints.

From an estimation perspective, longer $L$ improves the accuracy of the test statistic. The variance of the Rao test is inversely proportional to $L$ and can be approximated by:

\begin{equation}
    \mathrm{Var}(T_{3,\text{Rao}}) \propto \frac{1}{L}\left(1 + \frac{M}{K}\right).
\end{equation}

Moreover, for the Rao test statistic to be reliably computed, the sample covariance matrix $\boldsymbol{X}\boldsymbol{X}^H$ must be full-rank. This imposes a minimum observation length:

\begin{equation}
    L \geq M + K,
\end{equation}

which ensures the stability of the estimator. In the presence of multipath or asynchronous delays, this requirement is further tightened to:

\begin{equation}
    L \geq M + K + f_s(\max_i \tau_i - \min_i \tau_i),
\end{equation}

where $f_s$ is the sampling frequency and $\tau_i$ is the delay associated with the $i$-th device.

In addition to improving accuracy, $L$ helps compensate for asynchronous transmissions. Specifically, the system can resolve timing offsets up to a maximum delay $\Delta\tau_{\text{max}} = \frac{L - (M + K)}{f_s}$, which quantifies the maximum tolerable misalignment between devices. To achieve this, the guard interval must satisfy:

\begin{equation}
    L_g \geq f_s \Delta\tau,
\end{equation}

ensuring sufficient headroom for time misalignments.

However, a longer observation window is not always beneficial. There exists an energy-efficiency trade-off, as extending $L$ increases sampling and circuit energy consumption. The marginal improvement in detection probability with respect to $L$ is approximately given by:

\begin{equation}
    \frac{dP_{d}}{dL} \approx \frac{P_{\text{total}} N^2 \|\boldsymbol{s}\|^2}{2\sigma_n^2 K L^2} - \eta_{\text{circuit}},
\end{equation}

where $\eta_{\text{circuit}}$ models the power overhead per unit increase in $L$. This expression highlights that beyond a certain point, the gains in detection probability diminish and may be outweighed by hardware constraints.

Finally, several practical constraints limit the viable range of $L$. The observation window must remain within the channel coherence time, i.e., $L \leq f_s T_c$, to avoid temporal channel variations within a snapshot. Moreover, performance benefits taper off beyond $L > 2K(1 + M/N)$, where the additional samples provide limited marginal utility. Hardware buffer limitations may further restrict $L$ to satisfy $L \leq L_{\text{max}}$.

\vspace{0.5em}  
\noindent\textbf{3. Number of Devices ($K$):} The number of potentially active devices $K$ plays a central role in shaping detection performance. It directly affects the system's effective SNR due to power allocation and channel gain aggregation. As $K$ increases, each device receives less power under a fixed total power budget, leading to reduced per-device SNR and increased multiuser interference.

From our system model, the non-centrality parameter for Rao detection is given by
\begin{equation} \label{eq:noncent2}
    \lambda_{\text{Rao}} = \frac{2LM\|\boldsymbol{s}\|^2}{\sigma_n^2 K} \sum_{k=1}^K p_k \left( N^2 \alpha_{\text{IRS}} + \|\boldsymbol{d}_k\|^2 \right),
\end{equation}
where $p_k$ is the transmit power allocated to device $k$, and $\alpha_{\text{IRS}}$ characterizes the IRS gain. This expression reflects the combined effect of power allocation, IRS assistance, and direct channel quality.

We define an effective average SNR as
\begin{equation}
    \mathrm{SNR}_{\text{eff}} \triangleq \frac{1}{K} \sum_{k=1}^K \frac{p_k}{\sigma_n^2} \left( N^2 \alpha_{\text{IRS}} + \|\boldsymbol{d}_k\|^2 \right),
\end{equation}
so that the non-centrality parameter becomes
\begin{equation}
    \lambda_{\text{Rao}} = 2LM\|\boldsymbol{s}\|^2 \cdot \mathrm{SNR}_{\text{eff}}.
\end{equation}
To guarantee reliable detection, we impose the threshold condition $\lambda_{\text{Rao}} \geq \lambda_{\min}$, which translates to an SNR requirement:
\begin{equation} \label{eq:snr-cond}
    \mathrm{SNR}_{\text{eff}} \geq \frac{\lambda_{\min}}{2LM\|\boldsymbol{s}\|^2}.
\end{equation}

In power-constrained systems, where $P_{\text{total}}$ is shared across $K$ devices, power is often allocated proportionally to effective channel strength:
\begin{equation}
    p_k = \frac{P_{\text{total}} \|\boldsymbol{h}_k\|^2}{\sum_{k=1}^K \|\boldsymbol{h}_k\|^2},
\end{equation}
with $\boldsymbol{h}_k$ denoting the composite channel vector for device $k$. Substituting this into the SNR expression yields
\begin{equation}
    \mathrm{SNR}_{\text{eff}} = \frac{P_{\text{total}}}{K \sigma_n^2} \cdot \frac{\sum_{k=1}^K \|\boldsymbol{h}_k\|^2 \left( N^2 \alpha_{\text{IRS}} + \|\boldsymbol{d}_k\|^2 \right)}{\sum_{k=1}^K \|\boldsymbol{h}_k\|^2}.
\end{equation}

To meet the detection threshold, we derive the following upper bound on the number of devices:
\begin{equation}
    K \leq \frac{2LMP_{\text{total}} \|\boldsymbol{s}\|^2}{\lambda_{\min} \sigma_n^2} \cdot \frac{\sum_{k=1}^K \|\boldsymbol{h}_k\|^2 \left( N^2 \alpha_{\text{IRS}} + \|\boldsymbol{d}_k\|^2 \right)}{\sum_{k=1}^K \|\boldsymbol{h}_k\|^2}.
\end{equation}

Therefore, increasing $K$ enables the system to support more devices, but reduces the effective per-device SNR due to power sharing and interference. To ensure robust target detection in IRS-assisted systems, the number of supported devices must be chosen such that the effective SNR exceeds a minimum threshold. This SNR-centric view offers clearer system design insights and facilitates trade-off analysis between scalability and detection reliability.

\vspace{0.5em}
\noindent\textbf{4. IRS size and configuration ($N$):} The number of IRS elements $N$ and their phase configuration $\boldsymbol{\Phi} = \mathrm{diag}(e^{j\theta_1}, \ldots, e^{j\theta_N})$ directly influence detection performance through power enhancement and spatial focusing.

Under coherent phase alignment, where each IRS element aligns the phases of the incident and reflected channels, the IRS-induced channel gain exhibits approximate quadratic scaling:
\begin{equation}
\|\boldsymbol{F}\boldsymbol{\Phi}\boldsymbol{E}\|_F^2 \approx N^2 \alpha_{\text{eff}} + \mathcal{O}(N\sqrt{K}),
\end{equation}
where $\alpha_{\text{eff}} = \mathbb{E}[\|\boldsymbol{f}_{\text{avg}}\|^2 \|\boldsymbol{e}_k\|^2]$ captures the average per-element gain. This results in two key benefits: first, quadratic power scaling, corresponding to a $20\log_{10}N$ dB beamforming gain; and second, $N$ degrees of freedom to resolve device directions, enhancing spatial resolution and multi-user discrimination.

The minimum number of IRS elements required to overcome the noise floor is given by
\begin{equation}
N \geq \sqrt{\frac{K\sigma_n^2}{P_{\text{total}}\|\boldsymbol{s}\|^2}} \cdot \left(\max_k \|\boldsymbol{h}_k\|\right)^{-1} \left(1 + \frac{1}{\text{SNR}_{\text{min}}}\right),
\end{equation}
where $\text{SNR}_{\text{min}}$ denotes the minimum acceptable SNR for reliable detection. This condition ensures that the IRS can provide sufficient power focusing to meet system-level detection thresholds even in worst-case channel scenarios.

\vspace{0.5em}
\noindent\textbf{ 5. Transmit power allocation:} The allocation of transmit power across devices significantly impacts detection performance. The performance gap between optimal and equal power allocation is given by
\begin{equation}
\frac{T_{3,\text{Rao}}^{\text{opt}}}{T_{3,\text{Rao}}^{\text{eq}}} = 1 + \frac{\mathrm{Var}(\|\boldsymbol{h}_k\|^2)}{(\mathbb{E}\|\boldsymbol{h}_k\|^2)^2} + \mathcal{O}\left(\frac{M}{N^2}\right),
\end{equation}
where optimal allocation is defined as $p_k^* = \frac{P_{\text{total}} \|\boldsymbol{h}_k\|^2}{\sum_{k=1}^K \|\boldsymbol{h}_k\|^2}$ (maximizes $\lambda_{\text{Rao}}$ under total power constraint), and equal allocation as $p_k = P_{\text{total}} / K$ (blind operation).

The maximum performance gap arises when the channel gain variance is large, i.e., $\mathrm{Var}(\|\boldsymbol{h}_k\|^2) \gg (\mathbb{E}\|\boldsymbol{h}_k\|^2)^2$. In contrast, blind systems approach 90\% of optimal performance when $K \leq N^2$, due to channel hardening. Power imbalance remains tolerable when the ratio $\frac{\max_k \|\boldsymbol{h}_k\|^2}{\min_k \|\boldsymbol{h}_k\|^2}$ is less than $2$ (i.e., within 3 dB).

\section{Simulation Results and Discussions}\label{sec:simulation}
In this section, we present simulation results to evaluate the performance of the proposed detectors under various system parameters and validate the accuracy of the derived analytical expressions. Specifically, we analyze the probability of detection and false alarm for the four proposed detectors.  

To ensure a comprehensive evaluation, we perform Monte Carlo simulations with $10^5$ independent trials. The propagation channel between IoT devices and IRS follows Nakagami-$m$ fading with $m=2$. We assume the IRS has $N=16$ passive elements. For IRS phase values, we assume that the reflection coefficients are uniformly distributed over $[0, 2\pi]$. Regarding power settings, the transmit power of IoT devices is randomly assigned within the range of $[10, 50]$ mW to account for different device capabilities and energy constraints. The noise power is set to $-90$ dBm unless otherwise specified. The IRS reflection elements introduce an additional gain, modeled with an average amplitude reflection coefficient of $0.8$.  

The average SNR in the simulation is defined as:  
\begin{equation}\label{e103} 
\text{SNR} \triangleq \displaystyle{\frac{\parallel \mathbf{H} \mathbf{x} \parallel^2}{\text{tr}\left\{\mathbf{{\Sigma}_n}\right\}}},
\end{equation}  
where $\mathbf{H}$ is the cascaded effective channel matrix, including the effects of the IRS.

\begin{figure}[!t]
    \centering
    \includegraphics[width=\columnwidth]{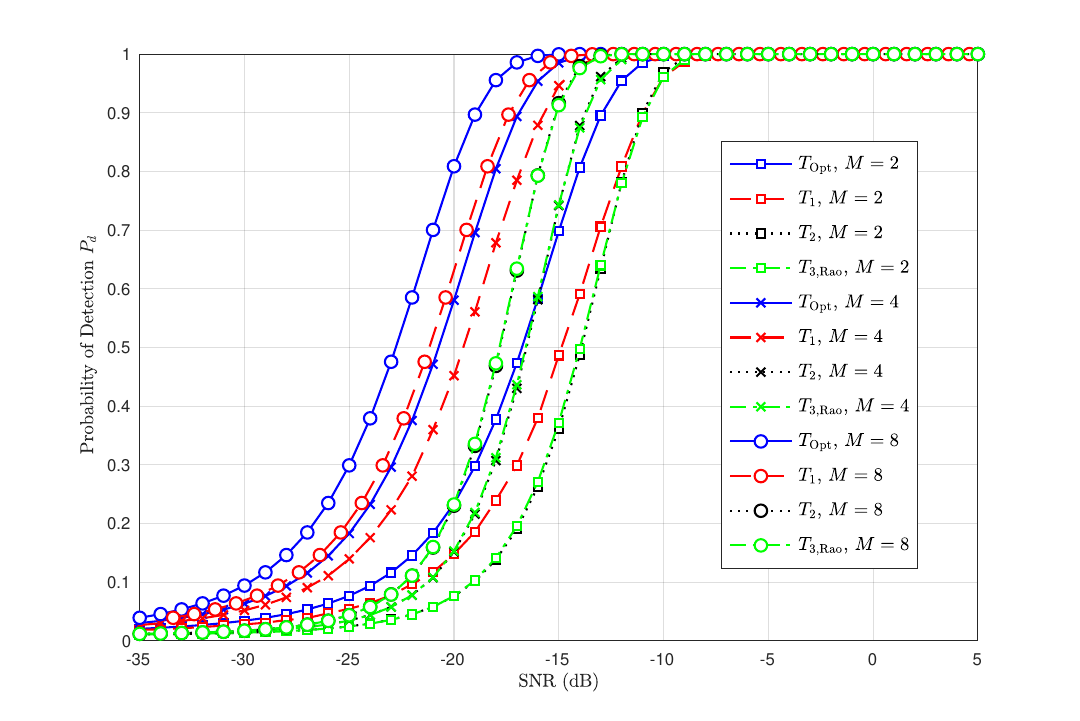}
    \caption{Probability of detection $P_d$ versus SNR for $K=6$, $P_{\mathrm{fa}} = 0.01$, $L = 16$, and varying number of AP antennas $M$.}
    \label{figM}
\end{figure}

Fig.~\ref{figM} quantifies the impact of the number of AP antennas $M$ on the probability of detection $P_d$. As expected from the theoretical spatial diversity gains in multiple antenna systems, increasing $M$ substantially enhances detection performance. Specifically, an SNR gain of approximately 5.48~dB is observed for the optimal detector $T_{\rm Opt}$ when $M$ increases from 2 to 8, which closely matches the theoretical 6~dB array gain. This confirms that the passive IRS architecture does not compromise the spatial resolution and diversity benefits traditionally offered by multiple antenna systems.

From a practical standpoint, the results indicate that at least $M = 4$ antennas are required to achieve reliable detection (i.e., $P_d > 0.9$) at low SNR values (e.g., below $-10$~dB). This is particularly relevant in low-power IoT applications, where maintaining high detection reliability under stringent energy constraints is essential to ensure long device lifetimes and efficient operation.

Moreover, the observed reduction in the performance gap between the optimal detector $T_{\rm Opt}$ and the suboptimal Rao-test-based detector $T_{3,\rm Rao}$ (from 1.4~dB at $M=2$ to 0.8~dB at $M=8$) reveals an important robustness feature. As $M$ increases, the suboptimal detector, which relies on partial or imperfect channel knowledge, increasingly benefits from the antenna diversity. This trend suggests that the sensitivity to channel estimation errors diminishes with larger antenna arrays, thereby reducing the system's dependency on perfect CSI. Such behavior is especially beneficial in dynamic or time-varying scenarios, where acquiring accurate CSI is either difficult or resource-intensive. \par

\begin{figure}[!t]
    \centering
    \includegraphics[width=\columnwidth]{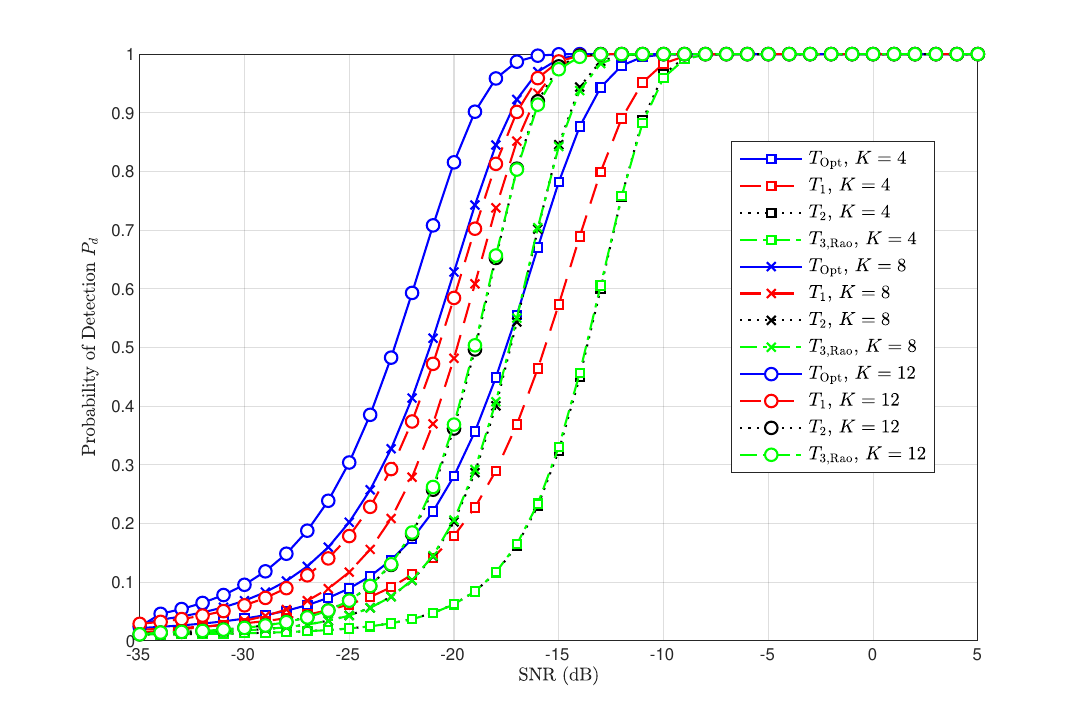}
    \caption{Probability of detection $P_d$ versus SNR for $M=4$, $P_{\rm fa} = 0.01$, $L = 16$, and different numbers of IoT devices $K$.}
    \label{figK}
\end{figure}

Fig.~\ref{figK} presents the impact of the number of IoT devices $K$ on the probability of detection $P_d$ for a fixed number of AP antennas $M = 4$. The results demonstrate that increasing $K$ enhances the detection performance due to the added energy diversity and increased degrees of freedom in the composite signal. Specifically, increasing $K$ from 4 to 12 yields a total SNR gain of approximately 1.74~dB for the optimal detector $T_{\rm Opt}$. However, this gain exhibits a diminishing return: while $K = 4 \to 8$ contributes approximately 0.9~dB improvement, the increment from $K = 8 \to 12$ provides only about 0.4~dB.

This sublinear gain behavior  implies that merely scaling the number of devices does not proportionally enhance detection, highlighting the importance of judicious scheduling or group-based transmission strategies in massive IoT environments.

Furthermore, the blind detector $T_{3,\rm Rao}$ maintains a consistently small performance gap staying within 1.2~dB of $T_{\rm Opt}$ across all values of $K$. This underscores the robustness of the Rao test in scenarios with partial or no channel CSI, and reinforces its applicability in large-scale IoT deployments where acquiring perfect CSI is infeasible due to latency, overhead, or energy constraints.

Thus, while increasing $K$ can improve detection reliability, its marginal benefit diminishes due to spatial saturation. Nevertheless, suboptimal detectors like $T_{3,\rm Rao}$ offer an efficient and practical alternative for real-world implementations involving dense IoT connectivity and limited CSI availability. \par

\begin{figure}[!t]
    \centering
    \includegraphics[width=\columnwidth]{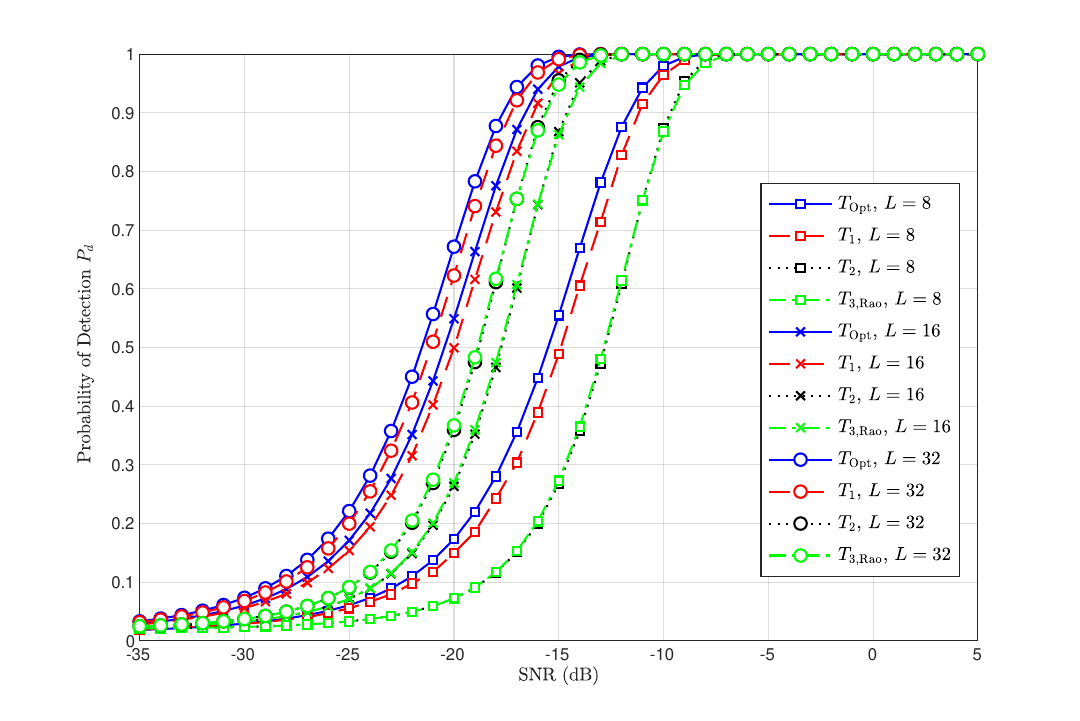}
    \caption{Probability of detection $P_d$ versus SNR for $M = 4$, $K = 6$, $P_{\rm fa} = 0.01$, and different numbers of temporal samples $L$.}
    \label{figL}
\end{figure}

Fig.~\ref{figL} characterizes the influence of the number of temporal samples $L$ on the detection performance, highlighting the inherent latency-reliability tradeoff in the system. The observed improvement of approximately 6~dB in $P_d$ when increasing $L$ from 8 to 32 confirms the theoretical prediction based on coherent integration, where the effective SNR scales proportionally with $\sqrt{L}$ under additive Gaussian noise assumptions.

This result underlines the crucial role of temporal sample accumulation in enhancing detection reliability. Specifically, when $L$ is small (e.g., $L = 8$), the detection performance degrades due to limited energy accumulation. However, even under such stringent latency constraints, the suboptimal detector $T_{3,\rm Rao}$ incurs only a 2.1~dB gap compared to the optimal detector $T_{\rm Opt}$ at $P_d = 0.9$, demonstrating its efficacy in scenarios where full CSI is unavailable.

From a system design perspective, adaptive selection of $L$ based on application-specific latency and reliability requirements is essential. For instance, longer integration windows (e.g., $L \geq 32$) are suitable for delay-tolerant use cases such as structural health monitoring or surveillance, where maximizing detection reliability is paramount. In contrast, shorter durations (e.g., $L \leq 16$) are preferable in time-critical applications like autonomous driving or real-time fault detection in industrial systems.

Moreover, the consistent 0.7~dB performance gap across all detectors at $L = 32$ suggests that the performance degradation of suboptimal detectors diminishes with increasing $L$. This observation reinforces the practical utility of blind or semi-blind detection techniques in high-sample regimes, where sufficient observations can compensate for the lack of perfect CSI. \par

\begin{figure}[!t]
    \centering
    \includegraphics[width=\columnwidth]{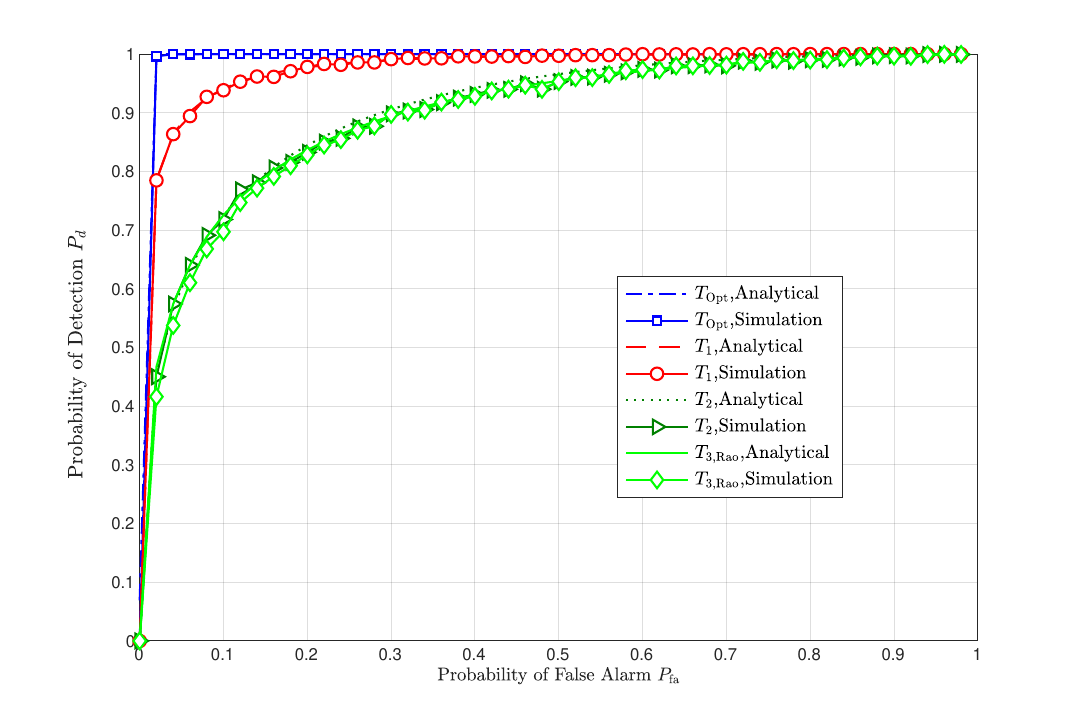}
    \caption{Comparison between simulation and analytical results for $M = 4$, $K = 6$, $L = 16$ at SNR = $-5$ dB.}
    \label{figAna}
\end{figure}

Fig.~\ref{figAna} confirms the accuracy of the derived theoretical expressions by comparing the analytical and simulation-based detection performance under practical system settings. With parameters $M = 4$, $K = 6$, $L = 16$, and SNR fixed at $-5$~dB, the analytical curves closely match the simulation results across a wide range of false alarm probabilities $P_{\rm fa}$. The maximum observed deviation is within 0.1\%, showcasing the robustness of the closed-form derivations even with moderate temporal sample sizes.

The tight agreement between simulation and analysis eliminates the need for computationally intensive Monte Carlo evaluations during system design and optimization. As such, the closed-form results provide a valuable tool for performance benchmarking, detector configuration, and adaptive thresholding in practical deployments.

In fact, Fig.~\ref{figAna} illustrates that the proposed analytical framework is not only mathematically sound but also practically reliable, enabling efficient and accurate prediction of detector behavior across a wide range of operating regimes. \par

\begin{figure}[!t]
    \centering
    
    \begin{minipage}{\columnwidth}
        \centering
        \includegraphics[width=\linewidth]{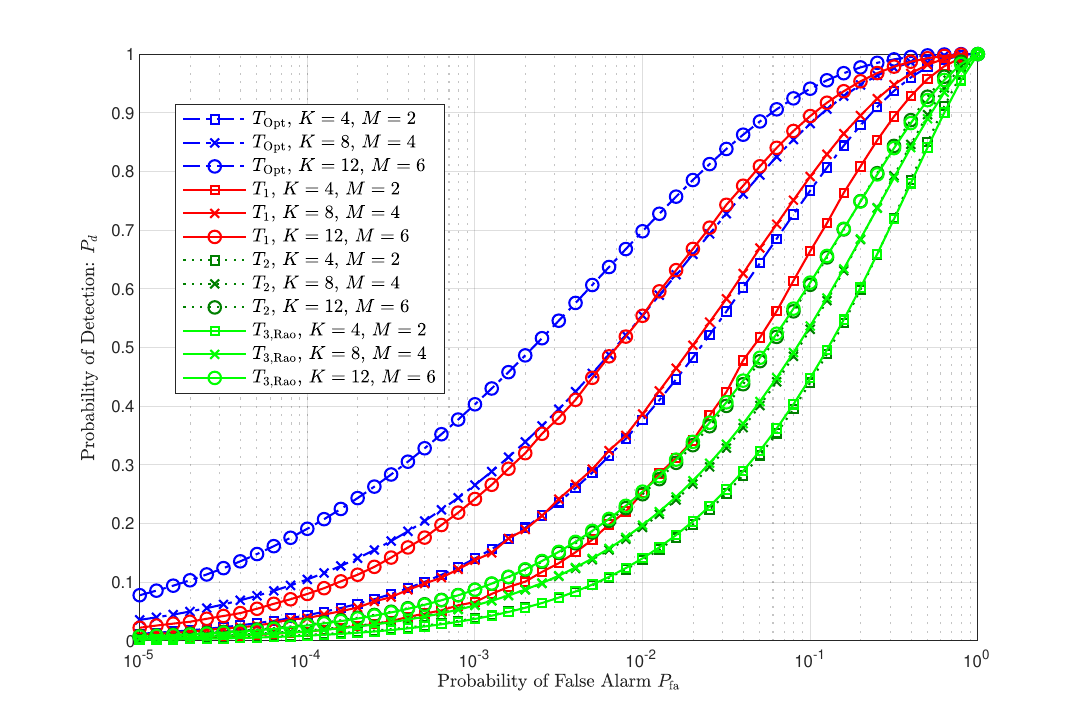} \\
        \vspace{2pt}
        \footnotesize (a)
    \end{minipage}
    
    \vspace{0pt} 
    
    \begin{minipage}{\columnwidth}
        \centering
        \includegraphics[width=\linewidth]{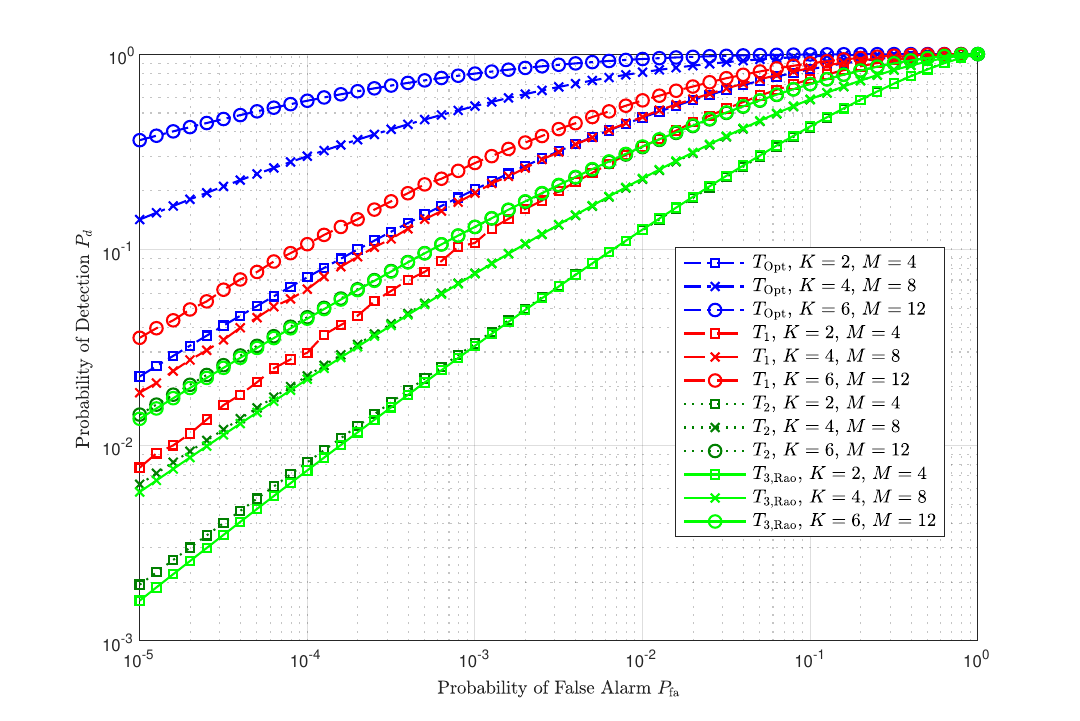} \\
        \vspace{2pt}
        \footnotesize (b)
    \end{minipage}
    
    \caption{ROC ($P_d$ vs. $P_{\rm fa}$) performance for SNR = $-5$ dB, and $L = 8$. 
    (a) Detection performance under increasing $(K, M)$ while maintaining a fixed ratio $K/M = 2$. 
    (b) Detection performance comparison for different values of $K$ and $M$ for a fixed ratio $K/M = 1/2$.}
    \label{figMK}
\end{figure}

Fig.~\ref{figMK} investigates the relative influence of the number of antennas $M$ versus the number of active IoT devices $K$ on the detection probability $P_d$ through receiver operating characteristic (ROC) curves. The two subplots highlight contrasting scaling regimes: one favoring user growth and the other prioritizing receiver antenna gain.

In Fig.~\ref{figMK}(a), the system operates under the constraint $K/M = 2$, and both $K$ and $M$ are increased proportionally. While such scaling theoretically improves spatial resolution, we observe diminishing gains in detection performance. For instance, increasing from $(K, M) = (4, 2)$ to $(8, 4)$ boosts $P_d$ at $P_{\rm fa} = 10^{-2}$ by about 3.5\%, but further increasing to $(12, 6)$ yields only a marginal 1.2\% improvement. This saturation behavior is primarily due to the fixed number of IRS elements ($N = 16$), which limits passive beamforming capability as user and antenna density increase. This highlights a key bottleneck: the IRS's scalability lags behind that of the transceiver components, curbing the expected benefits of joint scaling.

Conversely, Fig.~\ref{figMK}(b) fixes the ratio $K/M = 1/2$, favoring systems with more antennas than users. This overdetermined regime is inherently robust, as it improves the condition number of the system matrix, enhancing detection reliability. For example, reducing the number of users from $K = 8$ to $K = 4$ (while increasing $M$ from 4 to 8 to preserve the ratio) increases $P_d$ by over 8\% at $P_{\rm fa} = 10^{-3}$. This substantial gain supports theoretical insights that noise averaging and improved orthogonality in the received signal space dramatically enhance detection probability when the receiver is well-equipped.

These observations offer critical design insights. First, performance gains via increased $M$ are more sustainable than increasing $K$, particularly when IRS resources are limited. Second, the trade-off between $K$ and $M$ must be carefully balanced depending on latency, energy budget, and deployment constraints.  On the other hand, for dense device activity, matching $K$ and $M$ proportionally may be sufficient if the IRS size is also scaled accordingly.

\begin{figure}[!t]
    \centering
    \includegraphics[width=\columnwidth]{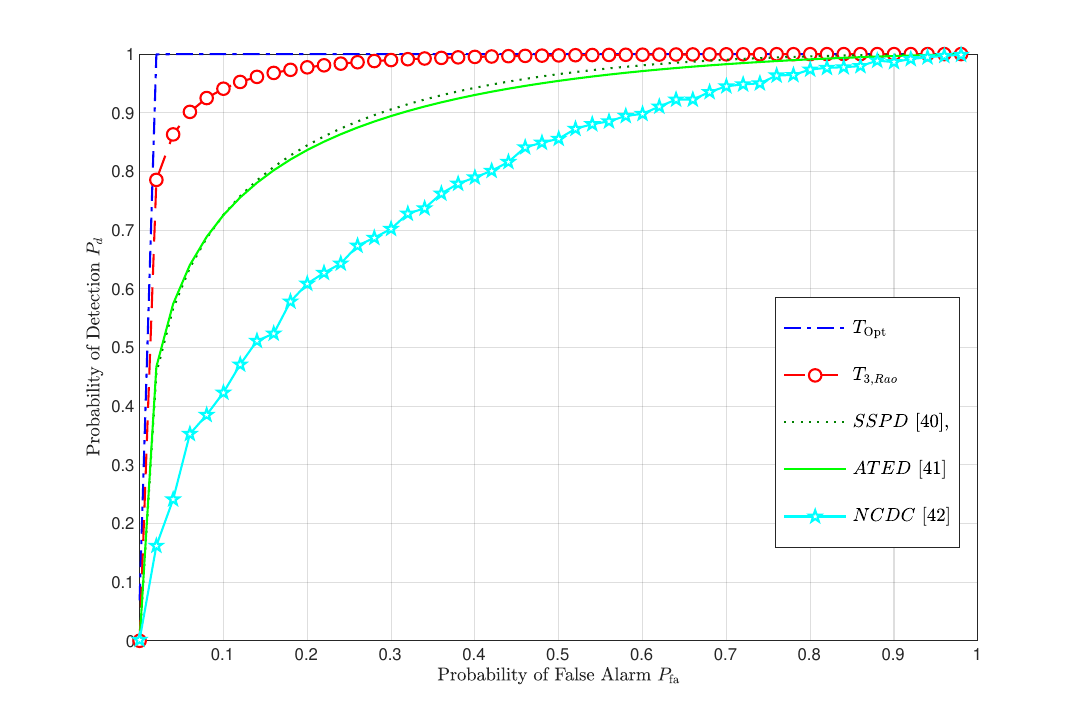}
    \caption{Complementary ROC comparison of the proposed blind detector with three benchmark detectors: SSPD~\cite{SSPD}, ATED~\cite{ATED}, and NCDC~\cite{NCDC}.}
    \label{figComp}
\end{figure}

Fig.~\ref{figComp} presents the complementary ROC curves comparing the proposed blind activity detector $T_{3,\text{Rao}}$ against three representative state-of-the-art schemes: SSPD~\cite{SSPD}, ATED~\cite{ATED}, and NCDC~\cite{NCDC}, along with the optimal detector $T_{\text{Opt}}$, which assumes complete knowledge of the channel matrix, noise variance, and transmit power levels. 

As anticipated, the optimal detector $T_{\text{Opt}}$ establishes an upper bound on performance, achieving near-perfect detection even at very low false alarm probabilities. Our proposed blind detector $T_{3,\text{Rao}}$, which operates without any prior knowledge of the system statistics or CSI, achieves performance remarkably close to $T_{\text{Opt}}$, particularly in the practical region of $P_{\rm fa} \leq 10^{-2}$. At $P_{\rm fa} = 10^{-3}$, for instance, $T_{3,\text{Rao}}$ incurs only a 1.5\% detection loss compared to the optimal detector, showcasing its robustness and practical applicability in unknown environments.

In contrast, all three benchmark detectors exhibit inferior performance. The SSPD and ATED methods perform similarly and moderately well, but lag behind in the low-$P_{\rm fa}$ region, where false detections can significantly degrade IoT system efficiency. The NCDC detector performs the worst, with a steep decline in detection accuracy as the false alarm rate decreases. This behavior reflects the limitations of simplified or non-coherent designs under asynchronous, multi-user, and IRS-assisted settings.

This figure demonstrates that the proposed blind detector not only significantly outperforms prior art but also offers reasonable performance with minimal assumptions, an essential property for future large-scale and low-cost IoT deployments where acquiring CSI and other system knowledge is impractical.

\begin{figure}[!t]
    \centering
    \includegraphics[width=\columnwidth]{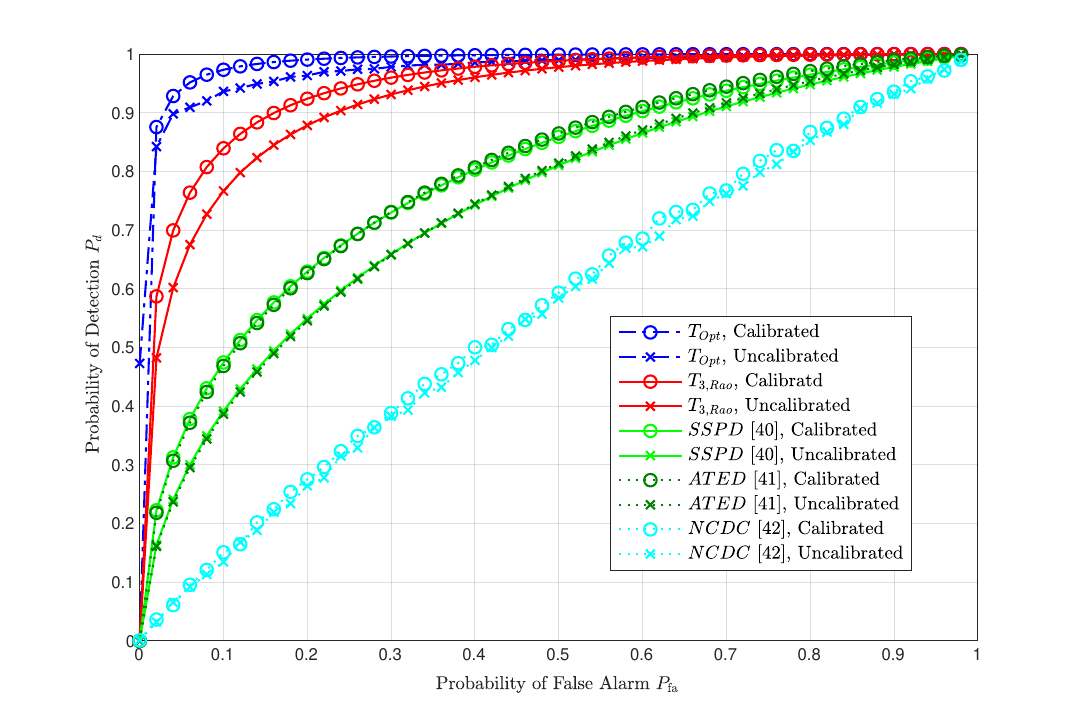}
 \caption{ROC curves comparing the performance of detectors ($T_{\text{Opt}}$, $T_{3,\text{Rao}}$, SSPD~\cite{SSPD}, ATED~\cite{ATED}, and NCDC~\cite{NCDC}) under unequal noise variance across AP antennas (uncalibrated case), with SNR = $-5$ dB, $K = 6$, $M = 4$, and $L = 8$.}
    \label{figCalib}
\end{figure}

Fig.~\ref{figCalib} shows the ROC curves comparing the performance of five detectors in a scenario where the noise variance $\bm{\Sigma}_n$ varies across the antennas of the AP. This scenario represents a typical real-world setting where noise power is not uniform due to imperfections such as manufacturing tolerances and poor calibration in RF systems. In these conditions, traditional detectors that assume uniform noise variance across antennas face significant degradation, typically on the order of 2-3 dB, due to their reliance on this unrealistic assumption.

As observed in Fig.~\ref{figCalib}, the detectors proposed in this work, particularly $T_{3,\text{Rao}}$, exhibit far less performance degradation, with the maximum variation being only 0.4 dB. This demonstrates the robustness of our approach, especially in the presence of uncalibrated noise variance. The blind nature of $T_{3,\text{Rao}}$, which does not require any prior knowledge of noise variance, contributes significantly to this resilience. By adapting to the actual channel conditions dynamically, it achieves a 0.2 dB improvement over the other methods, highlighting its superiority in scenarios where per-antenna calibration is not feasible or desirable.

In comparison, the baseline detectors such as SSPD and ATED are noticeably more sensitive to mismatches in noise variance. Their performance drops considerably in the presence of these imperfections, making them less reliable in practical, uncalibrated environments. The NCDC detector, already trailing in performance under ideal conditions, suffers even further with increased noise variance mismatch, emphasizing the inherent limitations of simplified or non-coherent detection schemes in such settings.

\begin{figure}[!t]
    \centering
    \includegraphics[width=\columnwidth]{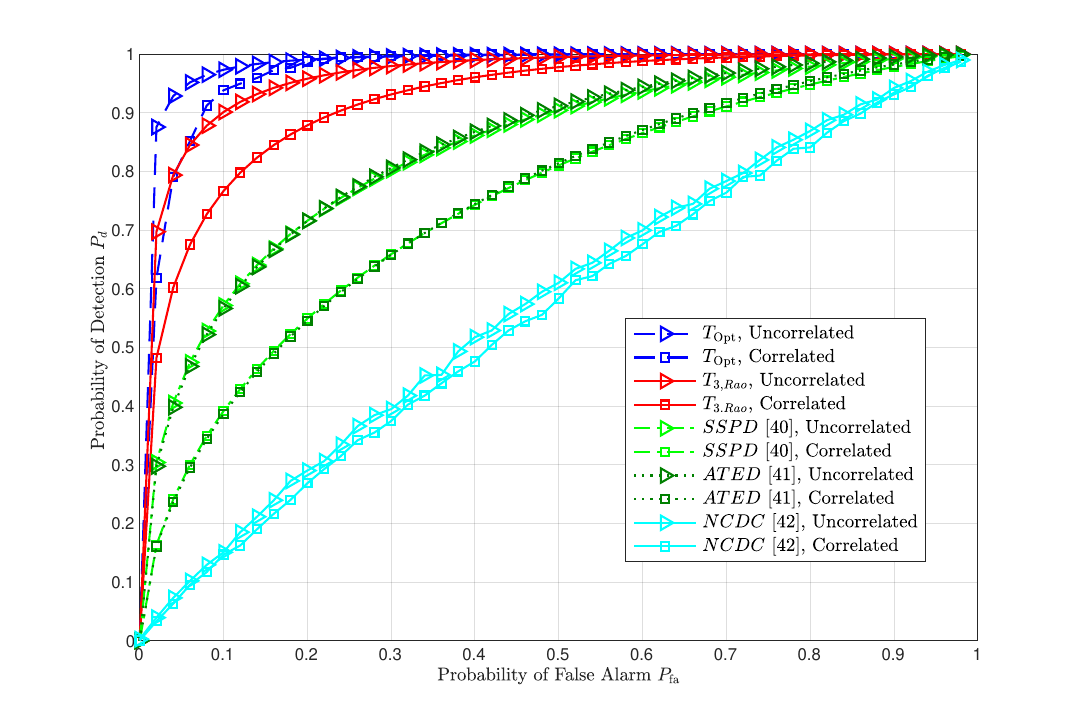}
  \caption{ROC curves comparing the performance of detectors ($T_{\text{Opt}}$, $T_{3,\text{Rao}}$, SSPD~\cite{SSPD}, ATED~\cite{ATED}, and NCDC~\cite{NCDC}) for correlated versus uncorrelated signals, with SNR = $-10$ dB, $K = 6$, $M = 4$, and $L = 32$.}
    \label{figCorr}
\end{figure}

Fig.~\ref{figCorr} examines the performance of the detectors in scenarios involving correlated and uncorrelated antennas. The spatial correlation among antenna elements is a critical factor in real-world signal propagation, especially in IRS-based networks. This correlation arises due to two primary factors: (1) limited scattering around the IRS and (2) the closely spaced configuration of antenna arrays. These factors can significantly impact detection performance if detectors are designed under the assumption of uncorrelated channels.

As seen in the figure, conventional detectors experience considerable performance degradation when spatial correlation is introduced, with performance losses of up to 7\%. This degradation is particularly pronounced when the detectors assume uncorrelated channels, demonstrating the vulnerability of these schemes in practical scenarios where moderate correlation is often present. The assumption of uncorrelated channels can lead to mismatches, which decrease the accuracy of detection and increase the false alarm rate.

In contrast, the proposed detectors, especially $T_{3,\text{Rao}}$, exhibit much less performance degradation under correlation uncertainty. The performance variation is limited to less than 3.2\%, showcasing the robustness of the blind detector in handling correlated channels. The key advantage of $T_{3,\text{Rao}}$ lies in its blind design, which does not require prior knowledge of the spatial correlation. This enables the detector to adapt to varying correlation conditions, ensuring reliable detection performance even when the correlation structure is unknown or fluctuating over time.

The results underscore the practical advantage of incorporating moderate spatial correlation in IRS-assisted systems, where values of $\rho$ in the range of 0.3-0.5 are often observed. This suggests that systems designed to handle such moderate correlation can maintain robust activity detection capabilities, even in environments with unknown or time-varying correlation. Additionally, the minimal performance degradation observed in $T_{3,\text{Rao}}$ emphasizes its suitability for deployment in dynamic and real-world environments where spatial correlation is uncertain or varies with time.
\section{Conclusion}\label{sec:conclusion}
This paper explores the detection methods for IoT activity detection under asynchronous transmission and heterogeneous power conditions assisted by IRS for potential NLoS scenarios. We propose and evaluate several detectors designed to address the challenges in such complex environments. The detectors are evaluated based on their ability to distinguish between the presence and absence of IoT activity, considering the dynamic nature of IoT networks, where devices may have different transmission powers and asynchronous communication patterns. Four detection methods are introduced, including a blind detector based on the Rao test, which is extended to handle nuisance parameters in practical settings. The performance of these detectors is thoroughly analyzed through simulation, where we compare their detection accuracy under varying conditions, such as mismatched noise variance, spatial correlation, and imperfect knowledge of system parameters. The results show that the proposed detectors, particularly the blind Rao detector, exhibit strong performance even in challenging conditions. The blind detector outperforms conventional methods and approaches the performance of an optimal detector, which assumes full knowledge of system parameters. These findings highlight the importance of robust detection strategies in real-world IoT networks, where system parameters may be uncertain or dynamically changing. In conclusion, this paper provides a comprehensive performance analysis of IoT activity detection under practical conditions, offering valuable insights into the design of effective detection systems for IoT applications.
\begin{appendices}

\section{Calculation of Fisher Information Matrix}\label{App-B}

First, we derive the derivative of the logarithm of the PDF under $\mathcal{H}_1$ in \eqref{eq15} with respect to $\boldsymbol{G}$. The derivative is computed separately for the real and imaginary parts of $\boldsymbol{G}$. We begin with:

\begingroup\makeatletter\def\f@size{8.6}\check@mathfonts
\begin{equation}\label{e78}
\frac{\partial\ln{f(\boldsymbol{X}|\mathcal{H}_1)}}{\partial\boldsymbol{G}} = -\frac{\partial}{\partial\boldsymbol{G}}\left\{\text{tr} \left\{\boldsymbol{\Sigma}_n^{-1}(\boldsymbol{X}-\boldsymbol{G}\boldsymbol{s}\boldsymbol{1}^T)(\boldsymbol{X}-\boldsymbol{G}\boldsymbol{s}\boldsymbol{1}^T)^H\right\}\right\}
\end{equation}
\endgroup

Next, we define the auxiliary variable $y$ and compute its derivatives: 
\begin{align}\label{e79} 
y &\triangleq \text{tr} \left\{\boldsymbol{\Sigma}_n^{-1}(\boldsymbol{X}-\boldsymbol{G}\boldsymbol{s}\boldsymbol{1}^T)(\boldsymbol{X}-\boldsymbol{G}\boldsymbol{s}\boldsymbol{1}^T)^H\right\} \nonumber \\
dy &= \text{tr} \left\{-\boldsymbol{s}\boldsymbol{1}^T(\boldsymbol{X}-\boldsymbol{G}\boldsymbol{s}\boldsymbol{1}^T)^H \boldsymbol{\Sigma}_n^{-1} d\boldsymbol{G} \right. \nonumber \\
&\quad \left. -\boldsymbol{\Sigma}_n^{-1}(\boldsymbol{X}-\boldsymbol{G}\boldsymbol{s}\boldsymbol{1}^T)\boldsymbol{1}\boldsymbol{s}^H d\boldsymbol{G}^H \right\}
\end{align}

The derivatives with respect to the real and imaginary parts of $\boldsymbol{G}$ are:
\begingroup\makeatletter\def\f@size{9}\check@mathfonts
\begin{align}\label{e80}
\left\{
\begin{array}{ll}
\displaystyle\frac{\partial y}{\partial \Re(\boldsymbol{G})} = -\boldsymbol{\Sigma}_n^{-T}(\boldsymbol{X}-\boldsymbol{G}\boldsymbol{s}\boldsymbol{1}^T)^*\boldsymbol{1}\boldsymbol{s}^T - \boldsymbol{\Sigma}_n^{-1}(\boldsymbol{X}-\boldsymbol{G}\boldsymbol{s}\boldsymbol{1}^T)\boldsymbol{1}\boldsymbol{s}^H \\\\
\displaystyle j\frac{\partial y}{\partial \Im(\boldsymbol{G})} = \boldsymbol{\Sigma}_n^{-T}(\boldsymbol{X}-\boldsymbol{G}\boldsymbol{s}\boldsymbol{1}^T)^*\boldsymbol{1}\boldsymbol{s}^T - \boldsymbol{\Sigma}_n^{-1}(\boldsymbol{X}-\boldsymbol{G}\boldsymbol{s}\boldsymbol{1}^T)\boldsymbol{1}\boldsymbol{s}^H
\end{array}
\right.
\end{align}
\endgroup

Thus, the derivative yields:
\begin{align}\label{e81} 
\frac{\partial\ln{f(\boldsymbol{X}|\mathcal{H}_1)}}{\partial\boldsymbol{G}} &= -\frac{1}{2}\left\{\frac{\partial y}{\partial \Re(\boldsymbol{G})} - j\frac{\partial y}{\partial \Im(\boldsymbol{G})}\right\} \nonumber \\
&= \boldsymbol{\Sigma}_n^{-T}(\boldsymbol{X}-\boldsymbol{G}\boldsymbol{s}\boldsymbol{1}^T)^*\boldsymbol{1}\boldsymbol{s}^T 
\end{align}

Similarly, the complex conjugate derivative is:
\begin{align}\label{e82} 
\frac{\partial\ln{f(\boldsymbol{X}|\mathcal{H}_1)}}{\partial\boldsymbol{G}^*} &= -\frac{1}{2}\left\{\frac{\partial y}{\partial \Re(\boldsymbol{G})} + j\frac{\partial y}{\partial \Im(\boldsymbol{G})}\right\} \nonumber \\
&= \boldsymbol{\Sigma}_n^{-1}(\boldsymbol{X}-\boldsymbol{H}\boldsymbol{s}\boldsymbol{1}^T)\boldsymbol{1}\boldsymbol{s}^H
\end{align}

Following the same approach, we derive the derivative with respect to the nuisance matrix $\boldsymbol{\Sigma}_n$:
\begin{align}\label{e83} 
\frac{\partial\ln{f(\boldsymbol{X}|\mathcal{H}_1)}}{\partial \boldsymbol{\Sigma}_n} &= \frac{\partial}{\partial\boldsymbol{\Sigma}_n}\bigg(-L \ln |\boldsymbol{\Sigma}_n| \nonumber \\
&\quad - \text{tr} \left\{\boldsymbol{\Sigma}_n^{-1}(\boldsymbol{X}-\boldsymbol{G}\boldsymbol{s}\boldsymbol{1}^T)(\boldsymbol{X}-\boldsymbol{G}\boldsymbol{s}\boldsymbol{1}^T)^H\right\}\bigg)
\end{align}

Define another auxiliary variable $\omega$:
\begin{align}\label{e84} 
\omega &\triangleq -L \ln |\boldsymbol{\Sigma}_n| - \text{tr} \left\{\boldsymbol{\Sigma}_n^{-1}(\boldsymbol{X}-\boldsymbol{H}\boldsymbol{s}\boldsymbol{1}^T)(\boldsymbol{X}-\boldsymbol{H}\boldsymbol{s}\boldsymbol{1}^T)^H\right\} \nonumber \\
d\omega &= -L \text{tr}\left\{\boldsymbol{\Sigma}_n^{-1} d\boldsymbol{\Sigma}_n\right\} \nonumber \\
&\quad - \text{tr} \left\{-\boldsymbol{\Sigma}_n^{-1} d\boldsymbol{\Sigma}_n \boldsymbol{\Sigma}_n^{-1} (\boldsymbol{X}-\boldsymbol{H}\boldsymbol{s}\boldsymbol{1}^T)(\boldsymbol{X}-\boldsymbol{H}\boldsymbol{s}\boldsymbol{1}^T)^H \right\}
\end{align}

From this, we obtain:
\begingroup\makeatletter\def\f@size{8.5}\check@mathfonts
\begin{align}\label{e85} 
\left\{ 
\begin{array}{ll}
\displaystyle\frac{\partial \omega}{\partial \Re(\boldsymbol{\Sigma}_n)} = -L\boldsymbol{\Sigma}_n^{-T} + \boldsymbol{\Sigma}_n^{-T}(\boldsymbol{X}-\boldsymbol{H}\boldsymbol{s}\boldsymbol{1}^T)^*(\boldsymbol{X}-\boldsymbol{H}\boldsymbol{s}\boldsymbol{1}^T)^T \boldsymbol{\Sigma}_n^{-T} \\\\
\displaystyle j\frac{\partial \omega}{\partial \Im(\boldsymbol{\Sigma}_n)} = L\boldsymbol{\Sigma}_n^{-T} - \boldsymbol{\Sigma}_n^{-T}(\boldsymbol{X}-\boldsymbol{H}\boldsymbol{s}\boldsymbol{1}^T)^*(\boldsymbol{X}-\boldsymbol{H}\boldsymbol{s}\boldsymbol{1}^T)^T \boldsymbol{\Sigma}_n^{-T}
\end{array}
\right.
\end{align}
\endgroup

Therefore, the derivatives are:
\begin{align}\label{e86} 
\frac{\partial\ln{f(\boldsymbol{X}|\mathcal{H}_1)}}{\partial \boldsymbol{\Sigma}_n} &= \frac{1}{2}\left(\frac{\partial \omega}{\partial \Re(\boldsymbol{\Sigma}_n)} - j\frac{\partial \omega}{\partial \Im(\boldsymbol{\Sigma}_n)}\right)  \\
&= -L\boldsymbol{\Sigma}_n^{-T} + \boldsymbol{\Sigma}_n^{-T}(\boldsymbol{X}-\boldsymbol{H}\boldsymbol{s}\boldsymbol{1}^T)^*(\boldsymbol{X}-\boldsymbol{H}\boldsymbol{s}\boldsymbol{1}^T)^T \boldsymbol{\Sigma}_n^{-T} \nonumber \\[5pt]
\frac{\partial\ln{f(\boldsymbol{X}|\mathcal{H}_1)}}{\partial \boldsymbol{\Sigma}_n^*} &= \frac{1}{2}\left(\frac{\partial \omega}{\partial \Re(\boldsymbol{\Sigma}_n)} + j\frac{\partial \omega}{\partial \Im(\boldsymbol{\Sigma}_n)}\right) = \boldsymbol{O}_{M \times M} \label{e87}
\end{align}

Substituting (\ref{e81}) and (\ref{e82}) into the second statement of (\ref{e24}), and (\ref{e86}) and (\ref{e87}) into the second statement of (\ref{e25}), yields the third statements. Throughout these calculations, we utilize properties of the $\text{vec}$ operator \cite{r41}. One key property, for matrices $\boldsymbol{A}_{m \times n}$, $\boldsymbol{B}_{n \times p}$, and $\boldsymbol{C}_{p \times r}$, is:
\begin{equation}\label{e88} 
\text{vec}\left\{\boldsymbol{A}\boldsymbol{B}\boldsymbol{C}\right\} = \left(\boldsymbol{C}^T \otimes \boldsymbol{A}\right) \text{vec}\left(\boldsymbol{B}\right)
\end{equation}

According to (\ref{e21}), $\boldsymbol{I}_{\boldsymbol{\Theta_r}\boldsymbol{\Theta_r}}$ is calculated as:
\begin{align}\label{e89} 
\boldsymbol{I_{\Theta_r \Theta_r}} &= \mathbb{E}\left\{\frac{\partial\ln{f(\boldsymbol{X}|\mathcal{H}_1,\boldsymbol{\Theta})}}{\partial\boldsymbol{\Theta_r}^*}\left(\frac{\partial\ln{f(\boldsymbol{X}|\mathcal{H}_1,\boldsymbol{\Theta})}}{\partial\boldsymbol{\Theta_r}^*}\right)^H\right\} \nonumber \\
&= \mathbb{E}\left\{
\begin{pmatrix}
\boldsymbol{b}\boldsymbol{b}^H & \boldsymbol{b}\boldsymbol{b}^T \\
\boldsymbol{b}^*\boldsymbol{b}^H & \boldsymbol{b}^*\boldsymbol{b}^T
\end{pmatrix}
\right\}
\end{align}

Under $\mathcal{H}_1$, the observation matrix $\boldsymbol{X}$ is proper, implying its semi-covariance matrix is zero \cite{r52}. Consequently, the off-diagonal entries of (\ref{e89}) vanish, giving:
\begin{align}\label{e90} 
\boldsymbol{I_{\Theta_r\Theta_r}} = 
\begin{pmatrix}
L\boldsymbol{s}^*\boldsymbol{s}^T \otimes \boldsymbol{\Sigma}_n^{-1} & \boldsymbol{O}_{MK \times MK} \\
\boldsymbol{O}_{MK \times MK} & L\boldsymbol{s}\boldsymbol{s}^H \otimes \boldsymbol{\Sigma}_n^{-T}
\end{pmatrix}
\end{align}

Similarly, the second block $\boldsymbol{I}_{\boldsymbol{\Theta_r \Theta_s}}$ is:
\begin{align}\label{e91} 
\boldsymbol{I_{\Theta_r\Theta_s}} &= \mathbb{E}\left\{\frac{\partial\ln{f(\boldsymbol{X}|\mathcal{H}_1,\boldsymbol{\Theta})}}{\partial\boldsymbol{\Theta_r}^*}\left(\frac{\partial\ln{f(\boldsymbol{X}|\mathcal{H}_1,\boldsymbol{\Theta})}}{\partial\boldsymbol{\Theta_s}^*}\right)^H\right\} \nonumber \\
&= \boldsymbol{O}_{2MK \times 2M^2}
\end{align}

In this derivation, (\ref{e33}) was employed.
\section{Derivation of the Correlation Coefficient Between the Numerator and Denominator of $T_{\text{SOpt1}}$}\label{App-C}

The numerator and denominator of $T_{\text{SOpt1}}$ can be expressed as functions $w = g(\boldsymbol{M}^H\boldsymbol{X})$ and $v = h(\boldsymbol{X}^H\boldsymbol{X})$, respectively. Under the null hypothesis $\mathcal{H}_0$, the independence between $\boldsymbol{M}^H\boldsymbol{X}$ and $\boldsymbol{X}^H \boldsymbol{X}$ leads to the independence between the numerator and the denominator.
\begin{align}\label{e92} 
&\mathbb{E}\left\{ \left(\boldsymbol{M}^H\boldsymbol{X}\right) \left(\boldsymbol{X}^H \boldsymbol{X}\right)\right\} \nonumber \\
&\quad =\boldsymbol{M}^H\mathbb{E}\left\{\boldsymbol{X}\right\}\mathbb{E}\left\{\boldsymbol{X}^H \boldsymbol{X}\right\}+\boldsymbol{M}^H\mathbb{E}\left\{\boldsymbol{X}\boldsymbol{X}^H \right\}\mathbb{E}\left\{ \boldsymbol{X}\right\} \nonumber \\
&\quad +\boldsymbol{M}^H\mathbb{E}\left\{\boldsymbol{X}\boldsymbol{X}\right\}\mathbb{E}\left\{ \boldsymbol{X}^H \right\} =0
\end{align}

In the above equation, the independence condition holds. Under $\mathcal{H}_1$, we directly investigate the independence between the two variables: $v = \operatorname{tr}\left\{\boldsymbol{X}^H\boldsymbol{X}\right\}$ and $w = \operatorname{tr}\left\{2\Re\left\{\boldsymbol{M}^H\boldsymbol{X}\right\} - \boldsymbol{M}^H\boldsymbol{M}\right\}$.
\begin{align}\label{e93} 
\mathbb{E}\left\{w v \right\} & =\mathbb{E}\left\{ \operatorname{tr}\left\{2\Re\left\{\boldsymbol{M}^H\boldsymbol{X}\right\}- 
\boldsymbol{M}^H\boldsymbol{M}\right\} \operatorname{tr}\left\{\boldsymbol{X}^H\boldsymbol{X}\right\}\right\} \nonumber \\
& =\underbrace{\mathbb{E}\left\{ \operatorname{vec}^H(\boldsymbol{M}) \operatorname{vec}(\boldsymbol{X})\operatorname{vec}^H(\boldsymbol{X})\operatorname{vec}(\boldsymbol{X})\right\}}_{(1)} \nonumber \\
& \quad +\underbrace{\mathbb{E}\left\{ \operatorname{vec}^H(\boldsymbol{X}) \operatorname{vec}(\boldsymbol{M})
\operatorname{vec}^H(\boldsymbol{X})\operatorname{vec}(\boldsymbol{X})\right\}}_{(2)} \nonumber \\
& \quad -\underbrace{\mathbb{E}\left\{ \operatorname{vec}^H(\boldsymbol{M}) \operatorname{vec}(\boldsymbol{M})\operatorname{vec}^H(\boldsymbol{X})\operatorname{vec}(\boldsymbol{X})\right\}}_{(3)}
\end{align}

That, we used as (\ref{e30}).

\begin{lem}\label{Th4}Janssen Theorem~\cite{r52}

Let $ \boldsymbol{A} $, $ \boldsymbol{B}$, $ \boldsymbol{C} $ and $ \boldsymbol{D} $ be matrices of dimension $p\times q $, $ q \times r $, $ r \times s $, and $ s \times t $. Assume that the entries of these matrices are random variables which jointly have a multivariate Gaussian distribution. Then the following result holds:
\begin{align}\label{e94} 
\mathds{E}\left\{\boldsymbol{ABCD}\right\}&=\sum_{k=1}^r \left(\mathds{E}\left\{\boldsymbol{e}_k^T\boldsymbol{C}\otimes\boldsymbol{A}\right\}
\mathds{E}\left\{\boldsymbol{D}\otimes\boldsymbol{B}\boldsymbol{e}_k\right\}\right)\nonumber \\
&+\mathds{E}\left\{\boldsymbol{AB}\right\}
\mathds{E}\left\{\boldsymbol{CD}\right\}+\mathds{E}\left\{\boldsymbol{A}\left(\mathds{E}\left\{\boldsymbol{BC}\right\}\right)\boldsymbol{D}\right\}
\nonumber \\
&-2\mathds{E}\left\{\boldsymbol{A}\right\}\mathds{E}\left\{\boldsymbol{B}\right\}
\mathds{E}\left\{\boldsymbol{C}\right\}\mathds{E}\left\{\boldsymbol{D}\right\}
\end{align}

Denote by $\boldsymbol{e}_k$ the vector having 1 at the $k$-th position and zeros elsewhere. For $r = 1$, the above expression simplifies to:
\begin{align}\label{e95} 
\mathds{E}\left\{\boldsymbol{ABCD}\right\}&=\mathds{E}\left\{\boldsymbol{C}\otimes\boldsymbol{A}\right\} 
\mathds{E}\left\{\boldsymbol{D}\otimes\boldsymbol{B}\right\}\nonumber \\
&+\mathds{E}\left\{\boldsymbol{AB}\right\}
\mathds{E}\left\{\boldsymbol{CD}\right\}+\mathds{E}\left\{\boldsymbol{A}\left(\mathds{E}\left\{\boldsymbol{BC}\right\}\right)\boldsymbol{D}\right\}
\nonumber \\
&-2\mathds{E}\left\{\boldsymbol{A}\right\}\mathds{E}\left\{\boldsymbol{B}\right\}
\mathds{E}\left\{\boldsymbol{C}\right\}\mathds{E}\left\{\boldsymbol{D}\right\}
\end{align}
\end{lem}

By applying this theory to part (1) of (\ref{e93}), we will have;
\begin{align}\label{e96} 
&\underbrace{\mathds{E}\left\{ vec^H(\boldsymbol{M}) vec(\boldsymbol{X})vec^H(\boldsymbol{X})vec(\boldsymbol{X})\right\}}_{(1)}\nonumber \\
&\quad =\mathds{E}\left\{vec^H(\boldsymbol{M}) vec(\boldsymbol{X})\right\}
\underbrace{\mathds{E}\left\{vec^H(\boldsymbol{X})vec(\boldsymbol{X})\right\}}_{a}\nonumber \\
&\quad +\mathds{E}\left\{vec^H(\boldsymbol{X})\otimes vec^H(\boldsymbol{M})\right\}
\underbrace{\mathds{E}\left\{vec(\boldsymbol{X})\otimes vec(\boldsymbol{X})\right\}}_{b}\nonumber \\
&\quad +\mathds{E}\left\{vec^H(\boldsymbol{M})\underbrace{\left[\mathds{E}
\left\{vec(\boldsymbol{X})vec^H(\boldsymbol{X})\right\}\right]}_{c}vec(\boldsymbol{X})\right\}\nonumber \\
&\quad -2\mathds{E}\left\{vec^H(\boldsymbol{M})\right\}\mathds{E}\left\{vec(\boldsymbol{X})\right\}
\mathds{E}\left\{vec^H(\boldsymbol{X})\right\}\mathds{E}\left\{vec(\boldsymbol{X})\right\}\nonumber\\
\end{align}

Each of the specified parts is calculated, separately. Under  $ \mathcal{H}_1 $, part $(a)$ is equivalent to the variance of (\ref{e42}).
\begin{equation}\label{e97} 
\mathds{E}\left\{vec^H(\boldsymbol{X})vec(\boldsymbol{X})\right\}=LM\sigma ^2+tr\left\{\boldsymbol{M}\boldsymbol{M}^H\right\}
\end{equation}

According to (\ref{e31}), for part $ (c)$, we have;
\begin{align}\label{e98} 
\mathds{E}\left\{vec(\boldsymbol{X})vec^H(\boldsymbol{X})\right\}=\sigma ^2\boldsymbol{I}_{ML}+vec(\boldsymbol{M})vec^H(\boldsymbol{M})
\end{align}

For any vector, the following expression holds;
\begin{equation}\label{e99}
vec(\boldsymbol{x}\boldsymbol{y}^H)=\boldsymbol{y}^*\otimes \boldsymbol{x}
\end{equation}

So, part $(b)$ turns into;
\begin{align} \label{e100}
&\mathds{E}\left\{vec(\boldsymbol{X})\otimes vec(\boldsymbol{X})\right\}
=\mathds{E}\left\{vec\left\{vec(\boldsymbol{X}) vec ^T(\boldsymbol{X})\right\}\right\}\nonumber \\
& \quad =vec\left\{SemiCov(vec(\boldsymbol{X}))
+\mathds{E}\left\{vec(\boldsymbol{X})\right\}\mathds{E}^T\left\{vec(\boldsymbol{X})\right\}\right\}
\nonumber \\
&\quad =vec(\boldsymbol{M})\otimes vec(\boldsymbol{M})
\end{align}

By substituting (\ref{e97}), (\ref{e98}) and (\ref{e100}) into (\ref{e96}), we will have;
\begin{align}\label{e101} 
&\underbrace{\mathds{E}\left\{ vec^H(\boldsymbol{M}) vec(\boldsymbol{X})vec^H(\boldsymbol{X})vec(\boldsymbol{X})\right\}}_{(1)}\nonumber \\ 
&\qquad\quad = vec^H(\boldsymbol{M}) vec(\boldsymbol{M})
\left(LM\sigma ^2+tr\left\{\boldsymbol{M}\boldsymbol{M}^H\right\}\right)\nonumber \\
&\qquad\quad +\left(vec^H(\boldsymbol{M})\otimes vec^H(\boldsymbol{M})\right)
\left(vec(\boldsymbol{M})\otimes vec(\boldsymbol{M})\right)\nonumber \\
&\qquad\quad +vec^H(\boldsymbol{M})\left[\sigma ^2\boldsymbol{I}_{ML}+vec(\boldsymbol{M})vec^H(\boldsymbol{M})\right] vec(\boldsymbol{M})\nonumber \\
&\qquad\quad -2vec^H(\boldsymbol{M})vec(\boldsymbol{M})
vec^H(\boldsymbol{M})vec(\boldsymbol{M})\nonumber \\
&=tr\left\{\boldsymbol{M}\boldsymbol{M}^H\right\}\left(LM\sigma ^2+tr\left\{\boldsymbol{M}\boldsymbol{M}^H\right\}\right)+\sigma ^2tr\left\{\boldsymbol{M}\boldsymbol{M}^H\right\}\nonumber \\
\end{align}

The same consequent with (\ref{e101}) for part (2) will derive. Part (3) is equivalent to (\ref{e97}). By using the above calculations, (\ref{e93}) can be obtained as following;
\begin{align}\label{e102} 
\mathds{E}\left\{w v \right\}=tr\left\{\boldsymbol{M}\boldsymbol{M}^H\right\}\left(LM\sigma ^2+tr\left\{\boldsymbol{M}\boldsymbol{M}^H\right\}+2\sigma ^2\right) \nonumber\\
\end{align}

Therefore, the numerator and the denominator variables are dependent, under  $\mathcal{H}_1 $. Finally, the correlation coefficient between the numerator and the denominator of $ T_{\rm Opt} $ is calculated as (\ref{e49}).


\end{appendices}
\bibliographystyle{IEEEtran} 

\end{document}